\documentclass[aps,prd,groupedaddress]{revtex4}
\usepackage{graphicx}
\usepackage{epsfig}
\usepackage{amsfonts}
\usepackage{amssymb}
\usepackage{epsf}
\usepackage{color}
\newcommand{\insertplot}[5]{\begin{figure}
 \hfill\hbox to 0.05in{\vbox to #5in{\vfill
 \inputplot{#1}{#4}{#5}}\hfill}
 \hfill\vspace{-.1in}
 \caption{#2}\label{#3}
 \end{figure}}
\newcommand{\inputplot}[3]{
 \special{ps: plotfile #1}
}

\newcounter{fig}

\begin{document}

\title{Wormholes in Einstein-scalar-Gauss-Bonnet theories \\ with a scalar self-interaction potential}
\author{Rustam Ibadov}
\email[]{ibrustam@mail.ru}
\affiliation{Department of Theoretical Physics and Computer Science, Samarkand State University,
Samarkand 140104, Uzbekistan}
\author{Burkhard Kleihaus}
\email[]{b.kleihaus@uni-oldenburg.de}
\affiliation{Institute of Physics, University of Oldenburg, D-26111 Oldenburg, Germany}
\author{Jutta Kunz}
\email[]{jutta.kunz@uni-oldenburg.de}
\affiliation{Institute of Physics, University of Oldenburg, D-26111 Oldenburg, Germany}
\author{Sardor Murodov}
\email[]{mursardor@mail.ru}
\affiliation{Department of Theoretical Physics and Computer Science, Samarkand State University,
Samarkand 140104, Uzbekistan$ $}

\date{\today}
\begin{abstract}
We construct wormholes in Einstein-scalar-Gauss-Bonnet theories
with a potential for the scalar field that includes a mass term
and self-interaction terms.
By varying the Gauss-Bonnet coupling constant we delimit the
domain of existence of wormholes in these theories.
The presence of the self-interaction enlarges the domain
of existence significantly.
There arise wormholes with a single throat and wormholes
with an equator and a double throat.
We determine the physical properties of these wormholes
including their mass, their size and their geometry.
\end{abstract}

\maketitle

\section{Introduction}

In General Relativity (GR) traversable Lorentzian wormholes need the
presence of some exotic form of matter, since the existence of
such wormholes requires the energy conditions to be violated
(see e.g.,~\cite{Morris:1988cz,Visser:1995cc,Lobo:2017eum}.)
In contrast, alternative theories of gravity can circumvent this feature.
Here the energy conditions can be violated due to the gravitational interaction itself.
Indeed, alternative theories of gravity can give rise to an effective stress energy tensor
that leads to violation of the energy conditions without the need for 
exotic matter (see e.g.,
\cite{Hochberg:1990is,Fukutaka:1989zb,Ghoroku:1992tz,Furey:2004rq,Bronnikov:2009az,Kanti:2011jz,Kanti:2011yv,Lobo:2009ip,Harko:2013yb}).

A particularly attractive type of theories, where traversable  Lorentzian wormholes
arise, are the Einstein-scalar-Gauss-Bonnet (EsGB) theories.
These theories contain higher curvature terms in the form of the Gauss-Bonnet (GB)
invariant, which arise for instance in string theories \cite{Zwiebach:1985uq,Gross:1986mw,Metsaev:1987zx}.
In order to contribute to the equations of motions in four spacetime dimensions,
the GB term should be coupled to a scalar field. String theories involve a dilaton field
and prescribe an exponential coupling of the dilaton field with the GB term.
However, in recent years, more general coupling functions have been suggested
\cite{Sotiriou:2013qea,Sotiriou:2014pfa,Antoniou:2017acq,Doneva:2017bvd,Silva:2017uqg}.
Among the attractive features of EsGB theories is the observation
that they lead to equations of motion, that are of second order,
and thus avoid the Ostrogradski instability and ghosts 
\cite{Horndeski:1974wa,Charmousis:2011bf,Kobayashi:2011nu}.

Already some time ago traversable wormholes were constructed in dilatonic EsGB theories 
\cite{Kanti:2011jz,Kanti:2011yv}. These wormholes possess a single throat at the center,
connecting two asymptotically flat regions. At the throat a thin shell of ordinary matter,
thus matter respecting the energy conditions, is localized. For a given equation of state,
this thin shell is determined by the Israel junction conditions \cite{Israel:1966rt,Davis:2002gn}.
The latter are invoked at the throat in order to obtain regular wormhole spactimes
without curvature singularities in both asymptotically flat regions. 

The boundaries of the domain of existence of these wormholes 
were shown to consist of the set of
dilatonic EsGB black holes \cite{Kanti:1995vq}, of a set of solutions with curvature singularities
and of a set of solutions with coordinate singularities \cite{Antoniou:2019awm}.
In fact, it was realized only recently, that at this latter boundary
the throat becomes degenerate and a new type of EsGB wormhole solutions arises.
These wormholes possess an equator at their center, that is surrounded by 
a double throat \cite{Antoniou:2019awm}.

The consideration of more general coupling functions of the scalar field to the GB term
brought forward the interesting new phenomenon of curvature induced spontaneous
scalarization of black holes
\cite{Antoniou:2017acq,Doneva:2017bvd,Silva:2017uqg,Antoniou:2017hxj,Blazquez-Salcedo:2018jnn,Doneva:2018rou,Minamitsuji:2018xde,Silva:2018qhn,Brihaye:2018grv,
Myung:2018jvi,Bakopoulos:2018nui,Doneva:2019vuh,Myung:2019wvb,Cunha:2019dwb,
Macedo:2019sem,Hod:2019pmb,Bakopoulos:2019tvc, Collodel:2019kkx,Bakopoulos:2020dfg,
Blazquez-Salcedo:2020rhf,Blazquez-Salcedo:2020caw}.
In fact, for an appropriate choice of coupling function, the GR black holes remain solutions of the
EsGB equations of motion. However, the GR black holes become unstable at  critical values of the 
GB coupling and develop scalar hair.
This is in contrast to the dilatonic EsGB theories with exponential coupling function,
which do not allow for GB black hole solutions
\cite{Kanti:1995vq,Torii:1996yi,Guo:2008hf,Pani:2009wy,Pani:2011gy,Kleihaus:2011tg,Ayzenberg:2013wua,Ayzenberg:2014aka,Maselli:2015tta,Kleihaus:2014lba,Kleihaus:2015aje,Blazquez-Salcedo:2016enn,Cunha:2016wzk,Zhang:2017unx,Blazquez-Salcedo:2017txk,Konoplya:2019hml,Zinhailo:2019rwd}.

The interesting properties of black holes in non-dilatonic EsGB theories have provoked the question
concerning the properties of wormholes in these theories.
Here a first study of such EsGB wormholes with a massless scalar field has already provided new insight
\cite{Antoniou:2019awm}.
In particular, numerous coupling functions were shown to possess such wormhole solutions,
and the domain of existence of wormhole solutions was fully mapped out for a quadratic coupling function.
Analogous to the dilatonic wormholes, the boundary of the domain of existence
is formed by EsGB black holes, by singular solutions and by wormholes with a degenerate throat,
where the latter can also be continued to wormhole solutions with an equator and a double throat.
Again, ordinary matter can be invoked to satisfy the junction conditions for 
a thin shell of matter at the single throat (or at the equator), yielding symmetric solutions.

Here we continue the investigation of EsGB wormholes by taking a type of coupling function,
that can give rise to a branch of stable fundamental EsGB black holes \cite{Doneva:2017bvd,Cunha:2019dwb}.
We further supplement the scalar field with a mass term and a self-interaction.
For the self-interaction we employ a potential of the form employed for non-topological solitons
and boson stars (see e.g.,~\cite{Friedberg:1986tq,Kleihaus:2005me}).
We construct the domain of existence of these wormholes and investigate their physical properties.
In particular, we consider the effects of the self-interaction as opposed to a mass term only.
We note that charged EsGB wormholes have also been obtained recently \cite{Brihaye:2020dgo}.

The paper is organized as follows: In section II we specify the theoretical setting, presenting the
action, the equations of motion, the boundary conditions, the conditions for throats (equators),
the junction conditions, and the energy conditions.
In section III we present our results, including the profile functions of the solutions,
the domain of existence and its boundaries, an analysis of the thin shell of matter at the throat (equator),
embeddings of the throat (equator) geometry, and the violation of the null energy condition (NEC).
In section IV we present our conclusions.

\section{Theoretical setting}

\subsection{Action and equations of motion}

We consider the effective action for Einstein-scalar-Gauss-Bonnet theories
\begin{eqnarray}  
S=\frac{1}{16 \pi}\int d^4x \sqrt{-g} \left[R - \frac{1}{2}
 \partial_\mu \phi \,\partial^\mu \phi - U(\phi)
 + F(\phi) R^2_{\rm GB}   \right],
\label{act}
\end{eqnarray} 
where $R$ is the curvature scalar,
$\phi$ is the scalar field with the coupling function $F(\phi)$ and 
potential $U(\phi)$,
and
\begin{eqnarray} 
R^2_{\rm GB} = R_{\mu\nu\rho\sigma} R^{\mu\nu\rho\sigma}
- 4 R_{\mu\nu} R^{\mu\nu} + R^2 
\end{eqnarray} 
is the quadratic Gauss-Bonnet correction term. 

The  Einstein equations and the scalar field equation are obtained
from the variation of the action with respect to the metric and the 
scalar field
\begin{eqnarray}
G^\mu_\nu & = & T^\mu_\nu \ , 
\label{Einsteq}\\
\nabla^\mu \nabla_\mu \phi & + & \dot{F}(\phi) R^2_{\rm GB}-\dot{U}(\phi)=0 \ .
\label{scleq}
\end{eqnarray}
The effective stress-energy tensor is given by the expression
\begin{equation}
T_{\mu\nu} =
-\frac{1}{4}g_{\mu\nu}\left(\partial_\rho \phi \partial^\rho \phi + 2 U(\phi) \right)
+\frac{1}{2} \partial_\mu \phi \partial_\nu \phi
-\frac{1}{2}\left(g_{\rho\mu}g_{\lambda\nu}+g_{\lambda\mu}g_{\rho\nu}\right)
\eta^{\kappa\lambda\alpha\beta}\tilde{R}^{\rho\gamma}_{\phantom{\rho\gamma}\alpha\beta}\nabla_\gamma \partial_\kappa F(\phi) \ .
\label{tmunu}
\end{equation}
Here, we have defined
$\tilde{R}^{\rho\gamma}_{\phantom{\rho\gamma}\alpha\beta}=\eta^{\rho\gamma\sigma\tau}
R_{\sigma\tau\alpha\beta}$ and $\eta^{\rho\gamma\sigma\tau}= 
\epsilon^{\rho\gamma\sigma\tau}/\sqrt{-g}$,
and the dot denotes the derivative with respect to the scalar field $\phi$.

To obtain static, spherically symmetric wormhole solutions we assume 
the line element in the form 
\begin{equation}
ds^2 = -e^{f_0} dt^2 +e^{f_1}\left[d\eta^2 
+h^2\left( d\theta^2+\sin^2\theta d\varphi^2\right) \right]\ ,
\label{met}
\end{equation}
with the auxiliary function $h^2 = \eta^2 +\eta_0^2$, where $\eta_0$ is a scaling parameter.
The two metric functions $f_0$ and $f_1$ and the scalar field function $\phi$ 
are functions of the radial coordinate $\eta$ only. 

Substitution of the above ansatz (\ref{met}) for the metric and the scalar field in
the Einstein equations and in the scalar-field equation leads to four coupled, nonlinear,
ordinary differential equations (ODEs), which are analogous to those
displayed in \cite{Antoniou:2019awm}, where, however, different coupling functions
$F(\phi)$ were chosen, and the scalar potential was set to zero.
Out of the four ODEs, 
three ODEs are of second order and one ODE is of first order.
But only three of the equations are independent.
In our numerical analysis we solve the first order and two of the second order ODEs.
%
We note that the field equations are invariant under the scaling transformation
\begin{equation}
\eta \to \chi \eta \ , \ \ \  \eta_0 \to \chi \eta_0 \ , \ \ \ F \to \chi^2 F\  , \ \ \ U \to \chi^{-2} U\ , \ \ \  \chi > 0 \ ,
\label{scalinvar}
\end{equation}
which allows to fix the parameter $\eta_0$.

\subsection{Throats, equators, and boundary conditions}

In order to obtain regular asymptotically flat wormhole solutions, we need to 
impose an appropriate set of boundary conditions for the ODEs.

Wormhole solutions possess one or more finite extrema of the circumferential 
radial coordinate 
\begin{equation}
R_C=e^{f_1/2}h \ .
\label{rc}
\end{equation}
In the simplest case they feature a single minimum corresponding to their single throat.
But they may also feature a local maximum surrounded by two minima.
The local maximum then corresponds to their equator, while the two minima
represent their two throats surrounding their equator.
While wormholes with more extrema do exist in other theories,
we do not find such wormholes in the theory considered here.

To obtain the inner boundary conditions we require the presence of an extremum 
of the circumferential radius at the center $\eta=0$.
This yields 
\begin{equation}
\left. \frac{dR_C}{d\eta} \right|_{\eta=0} = 0 \ \ \ 
\Longleftrightarrow \ \ \ \left.  f'_1\right|_{\eta=0} =0 \ .
\label{extr_rc}
\end{equation}
Thus we choose as one of our boundary conditions at the center 
the condition $\left.  f'_1\right|_{\eta=0} =0$. In addition, we choose 
at the center any one of the three conditions
\begin{equation}
\left.  f_0\right|_{\eta=0} = f_{0c}  \ , \ \ \ \left.  f_1\right|_{\eta=0} = f_{1c}  \ , \ \ \
\left.  \phi \right|_{\eta=0} = \phi_{c}  
 \ .
\label{extr_rc2}
\end{equation}

In order to obtain asymptotically flat solutions we employ the boundary conditions at infinity
\begin{equation}
\left. f_0\right|_{\eta=\infty}=0 \ , \ \ \ \left. f_1\right|_{\eta=\infty}=0 \ , \ \ \ \left. \phi\right|_{\eta=\infty}=0 \ .
\label{boundinfty}
\end{equation}
Consequently, we find for a fixed value of $\alpha$ a one parameter family of solutions.

Expansion of the functions at infinity shows, that we can read off the mass $M$ of the solutions as follows
\begin{equation}
f_1 = \frac{2M}{\eta} + O\left( \frac{1}{\eta^2} \right) \ .
\end{equation}
Since the solutions are symmetric, the mass has the same value
in both asymptotically flat parts of the spacetime.
We note that the wormhole solutions with a mass term in the potential 
have a vanishing scalar charge due to the exponential decay of the
scalar field.

\subsection{Junction conditions}

The solutions are symmetrically continued to the negative $\eta$ range,
yielding a second asymptotically flat region.
Since the derivatives of the functions $f_0$ and $\phi$ do not vanish
at the center $\eta=0$,  in general, we
amend the solutions by the presence of a thin shell-like distribution of matter there.
The complete
solution is then determined by invoking the junction conditions \cite{Israel:1966rt,Davis:2002gn}, 
at the jumps of the Einstein and scalar field equations at the throat (or equator) $\eta=0$.

The jumps in the Einstein and scalar field equations at the center are given by
\begin{equation}
\langle G^\mu_{\phantom{a}\nu} -T^\mu_{\phantom{a}\nu}\rangle = s^\mu_{\phantom{a}\nu} \ , \ \ \ 
\langle \nabla^2 \phi + \dot{F} R^2_{\rm GB}\rangle = s_{\rm s} \ ,
\label{jumps}
\end{equation}
where $s^\mu_\nu$ corresponds to the stress-energy tensor of the matter
at the center and $s_{\rm s}$ represents a source term for the scalar field.
Since we would like a shell of non-exotic matter,
we assume a perfect fluid with pressure $p$ and energy density $\varepsilon$, and
a scalar density $\rho_{\rm scal}$ at the center \cite{Kanti:2011jz,Kanti:2011yv,Antoniou:2019awm}
\begin{equation}
S_\Sigma = \int \left[\lambda_1 + 2 \lambda_0 F(\phi) \bar{R}\right]\sqrt{-\bar{h}} d^3 x  \ ,
\label{act_th}
\end{equation}
where we have introduced the constants $\lambda_1$ and $\lambda_0$,
and we denote by $\bar{h}_{ab}$ the three-dimensional
induced metric at the center and by $\bar{R}$ the corresponding Ricci scalar. 
This leads to the junction conditions
\begin{eqnarray}
8 \dot{F} \phi' e^{-\frac{3 f_1}{2}}
& = &
\lambda_1\eta_0^2 + 4\lambda_0 F  e^{-f_1}- \varepsilon\eta_0^2 \ , 
\label{j_00}\\
e^{-\frac{f_1}{2}} f_0' 
& = &
\lambda_1 + p  \ , 
\label{j_tt}\\
e^{-f_1}\phi' - 4  \frac{\dot{F}}{\eta_0^2} f_0' e^{-2 f_1}
& = &
-4\lambda_0\frac{\dot{F}}{\eta_0^2} e^{-\frac{3 f_1}{2}} +\frac{\rho_{\rm s}}{2} \ ,
\label{j_ss}
\end{eqnarray}
where all quantities are taken at $\eta=0$. 
The matter density $\varepsilon$ and pressure $p$, and the scalar density $\rho_{s}$
are determined via 
the arbitrary constants $\lambda_0$ and $\lambda_1$ and the functions close to the center.
We can always find a range of values for the constants
$\lambda_1$ and  $\lambda_0$, where the matter density $\varepsilon$ is positive, 
thus allowing us to avoid any exotic matter. In the case of vanishing pressure, i.e., for $p=0$, 
the matter simply corresponds to dust.
and therefore
its equation of state is the one of dust. In this special case we obtain
$\lambda_1=e^{-f1/2}f_0'$. Choosing $\lambda_0=\lambda_1$, in addition, we find
\cite{Kanti:2011jz,Kanti:2011yv,Antoniou:2019awm}
\begin{eqnarray}
    \varepsilon &=& \frac{e^{-\frac{3 f_1}{2}}}{\eta_0}\left[  \left(  4 F + \eta_0^2 e^{f_1} \right)f_0'
    -8\dot F \phi'     \right] \ , 
\label{rhomat1} \\
    \rho_{\rm s} &=& 2 e^{-f_1} \phi',
\label{rhomat2} 
\end{eqnarray}
where again all quantities are taken at $\eta=0$.

\subsection{Energy conditions}

In wormhole solutions the NEC
\begin{equation}
T_{\mu\nu} n^\mu n^\nu \geq 0 \ 
\end{equation}
must be violated, where
$n^\mu$ is any null vector ($n^\mu n_\mu=0$). 
Defining the null vector $n^\mu$
\begin{equation}
n^\mu=\left(1,\sqrt{-g_{tt}/g_{\eta\eta}},0,0\right) \ ,
\end{equation}
and thus $n_\mu=\left(g_{tt},\sqrt{-g_{tt}\,g_{\eta\eta}},0,0\right)$,
the NEC takes in a spherically symmetric spacetime the form
\begin{equation}
T_{\mu\nu}n^\mu n^\nu=T^t_t n^t n_t + T^\eta_\eta n^\eta n_\eta
=-g_{tt}\,(-T^t_t +T^\eta_\eta)   \ .
\end{equation}
Consequently the NEC holds when 
\begin{equation}
 -T_t^t + T_\eta^\eta \geq 0 \ .
\label{nec1}
\end{equation}
Alternatively, defining
\begin{equation}
n^\mu=\left(1,0,\sqrt{-g_{tt}/g_{\theta \theta}},0\right) \ ,
\end{equation}
the NEC holds when
\begin{equation}
-T_t^t + T_\theta^\theta \geq 0 \ . 
\label{nec2}
\end{equation}
For wormhole solutions these conditions
must be violated \cite{Morris:1988cz,Kanti:2011jz,Kanti:2011yv,Antoniou:2019awm}. 

\subsection{Embeddings}

To visualize the wormhole geometry we consider the isometric embedding of
the equatorial plane of the solutions.
The equatorial plane is obtained from the line element (\ref{met})
by setting $t$ constant, and $\theta=\pi/2$. 
This line element is then set equal to a hypersurface in the
three-dimensional Euclidean space, yielding
\begin{equation}
e^{f_1}\,[d\eta^2 +(\eta^2+\eta_0^2)\,d\varphi^2]=
d\rho^2 +\rho^2 d\varphi^2 + dz^2 \ , \label{emb}
\end{equation}
where ($\rho$, $\varphi$, $z$) represent cylindrical coordinates on the hypersurface.
We now consider $\rho$ and $z$ to be functions of $\eta$
\begin{eqnarray}
    \rho(\eta) &=&e^{f_1/2}\sqrt{\eta^2+\eta_0^2} \ ,
\label{rhodef} \\ 
    \left(\frac{d\rho }{d\eta}\right)^2+\left(\frac{dz}{d\eta}\right)^2  &=&e^{f_1} \ . 
\label{zzde}
\end{eqnarray}
Solving for $z(\eta)$ leads to
\begin{equation}
   z(\eta)=\pm \int_0^\eta \sqrt{e^{f_1(\tilde \eta)} -\left( \frac{d}{d\tilde \eta}
   \left[  e^{f_1(\tilde \eta)/2}\sqrt{\tilde \eta^2+\eta_0^2} \right]  \right)^2}d\tilde \eta \ .
 \label{zeq}
\end{equation} 
With $\rho(\eta)$ and $z(\eta)$ we thus obtain a  parametric representation 
of the equatorial plane (for a fixed value of the $\varphi$).

\section{Results}

\subsection{Parameters and numerics}

In the following we present our results, obtained with the coupling function $F(\phi)$ and the potential $U(\phi)$
\begin{equation}
F(\phi) = \frac{\alpha}{2\beta}\left(1-e^{-\beta \phi^2}\right) \ , \ \ \ 
U(\phi) = \lambda\left(c_2 \phi^2 +c_4 \phi^4 +c_6 \phi^6\right) \ .
\label{FandU}
\end{equation}
We fix the constant $\beta=1.5$, but leave $\alpha$ as a free parameter.
When employing the full potential
we also set $\lambda=0.06$, $c_2=1.1$, $c_4=-2$, and  $c_6=1$,
whereas in the case of a mass term only,
we retain $\lambda=0.06$, $c_2=1.1$, but set $c_4=c_6=0$.
Furthermore, we set $\eta_0=1$.

The numerical integration of the system of coupled ODEs is done with the help of COLSYS \cite{Ascher:1979iha}. 
COLSYS is an ODE solver that uses a collocation method for boundary value ODEs together with 
a damped Newton method of quasi-linearization. 
At each iteration step the linearized problem is solved via a spline collocation at Gaussian points. 
COLSYS employs a mesh selection procedure to adapt and refine the mesh
until a prescribed stopping criterion is reached.
We typically use  $10^3$ subintervals, yielding a relative error of less than $10^{-8}$ for the solutions.
Note however that close to singular solutions the number of subintervals is increased up to $5 \times 10^4$.

\subsection{Solutions}

\begin{figure}[t!]
\begin{center}
(a)\includegraphics[width=.45\textwidth, angle =0]{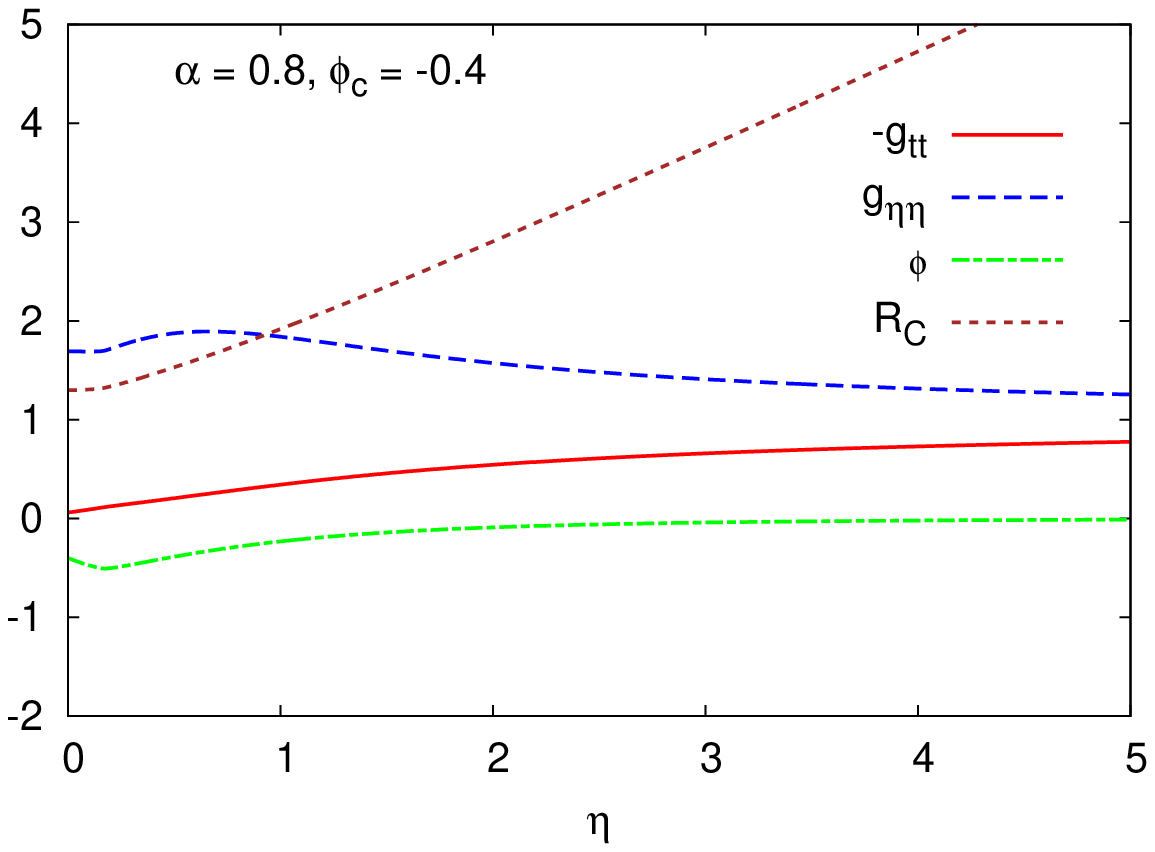}
(b)\includegraphics[width=.45\textwidth, angle =0]{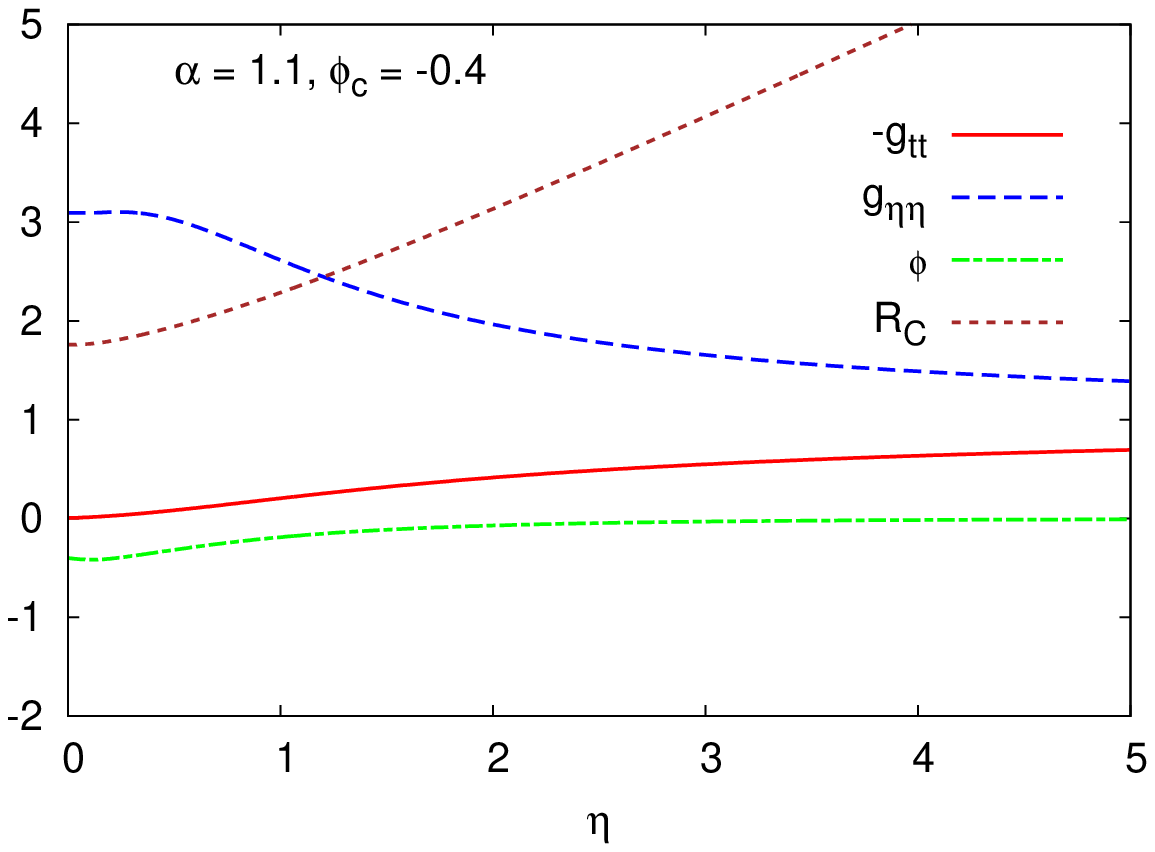}
\\
(c)\includegraphics[width=.45\textwidth, angle =0]{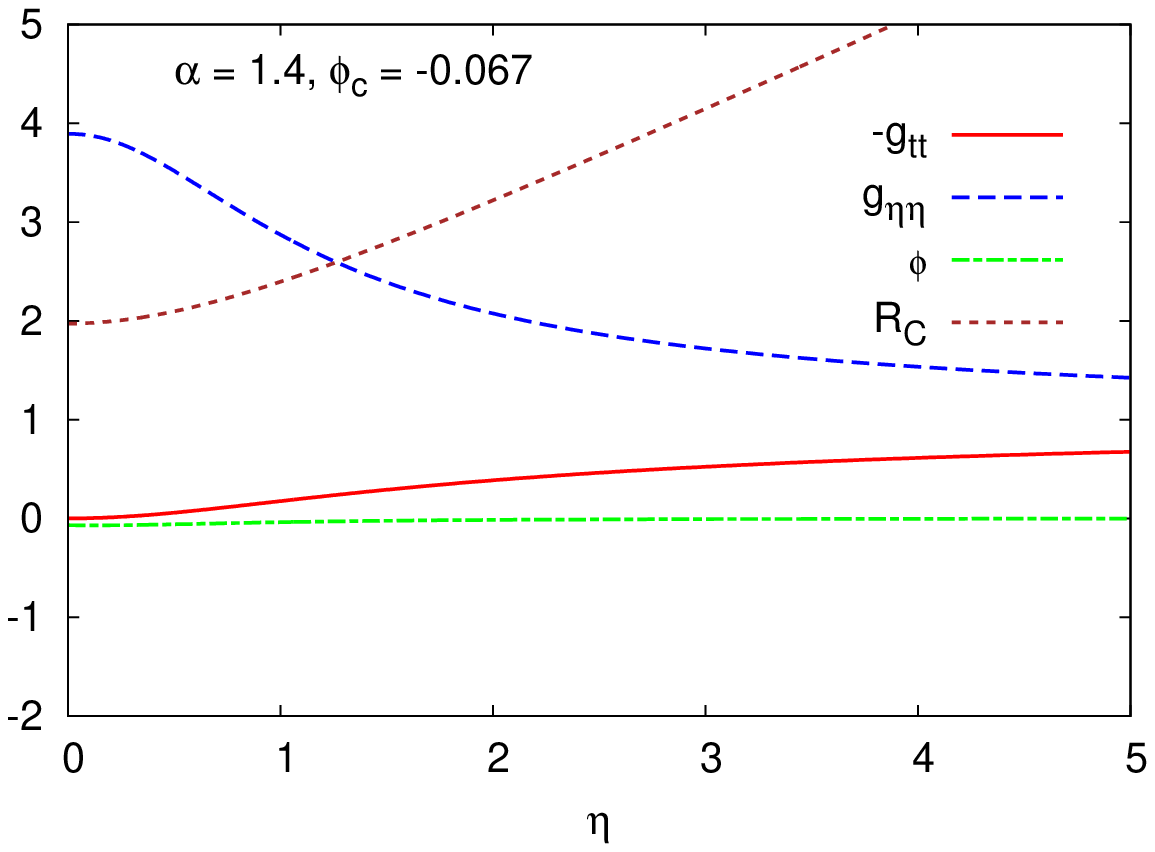}
(d)\includegraphics[width=.45\textwidth, angle =0]{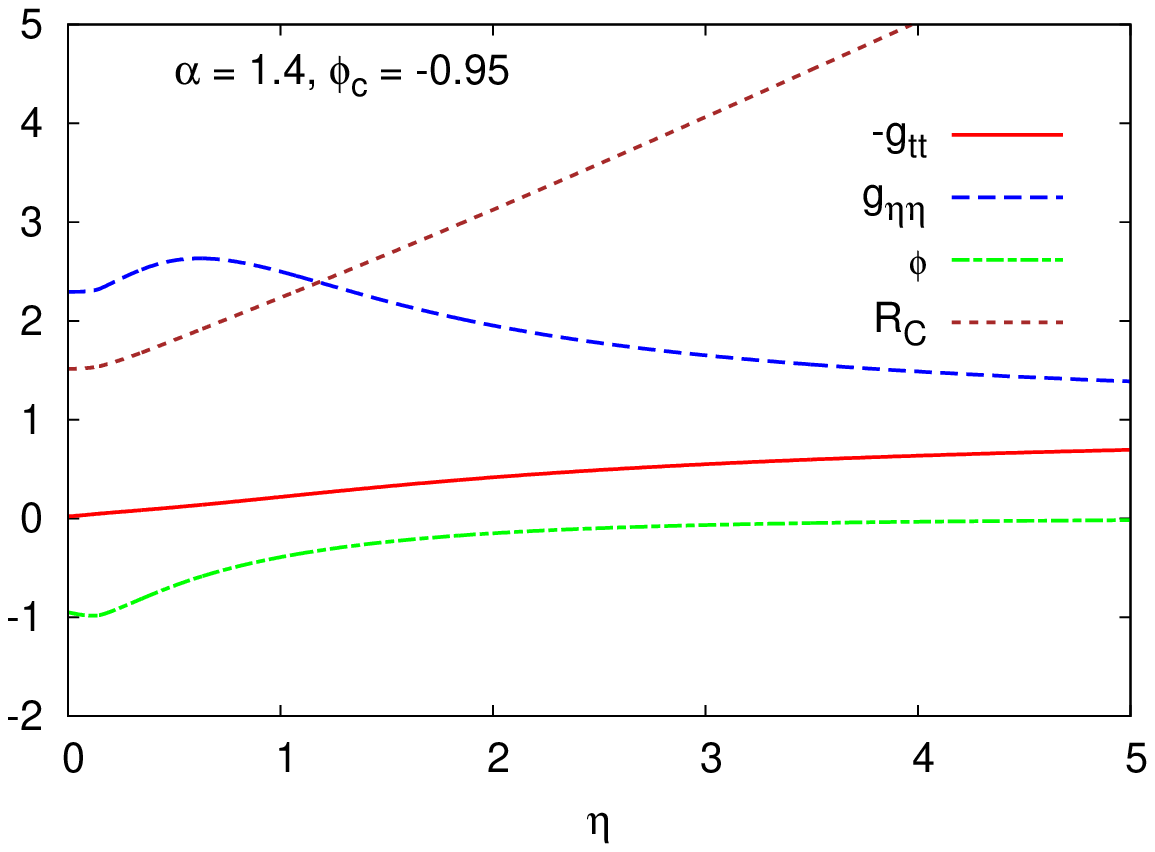}
\\
(e)\includegraphics[width=.45\textwidth, angle =0]{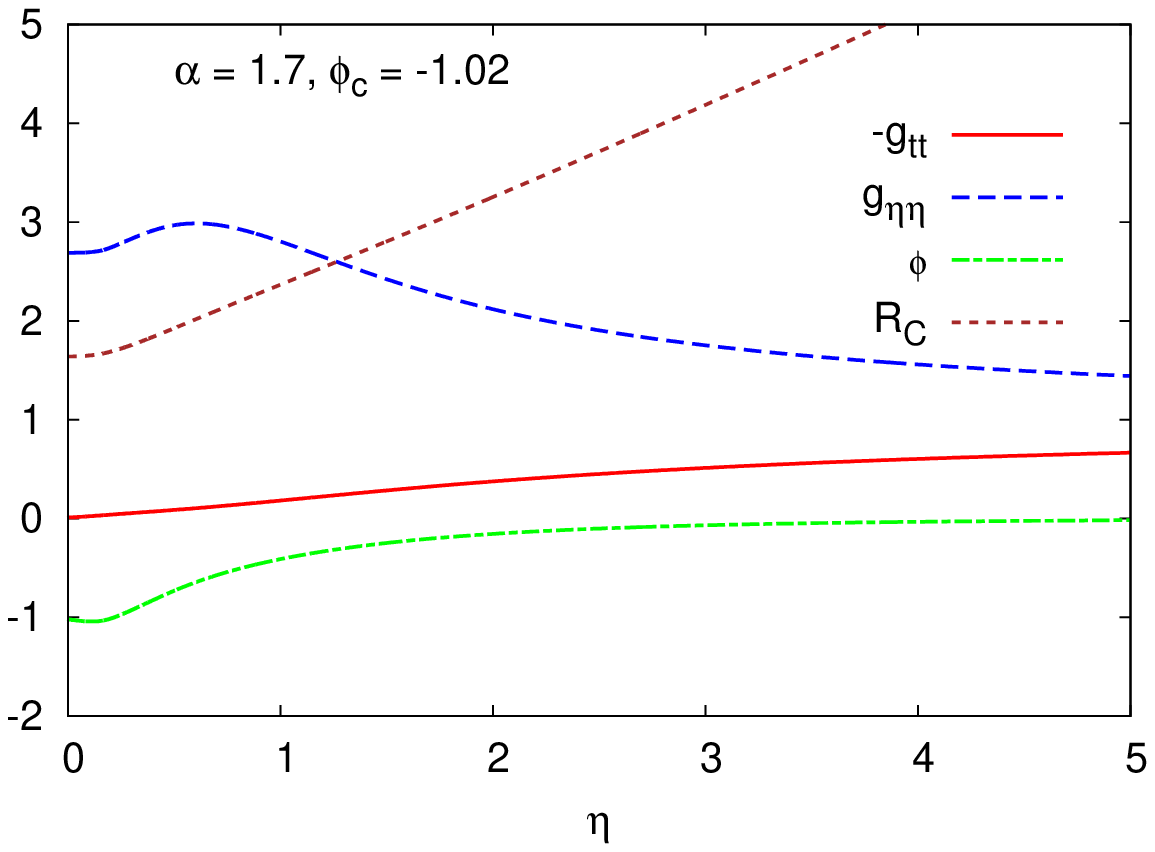}
(f)\includegraphics[width=.45\textwidth, angle =0]{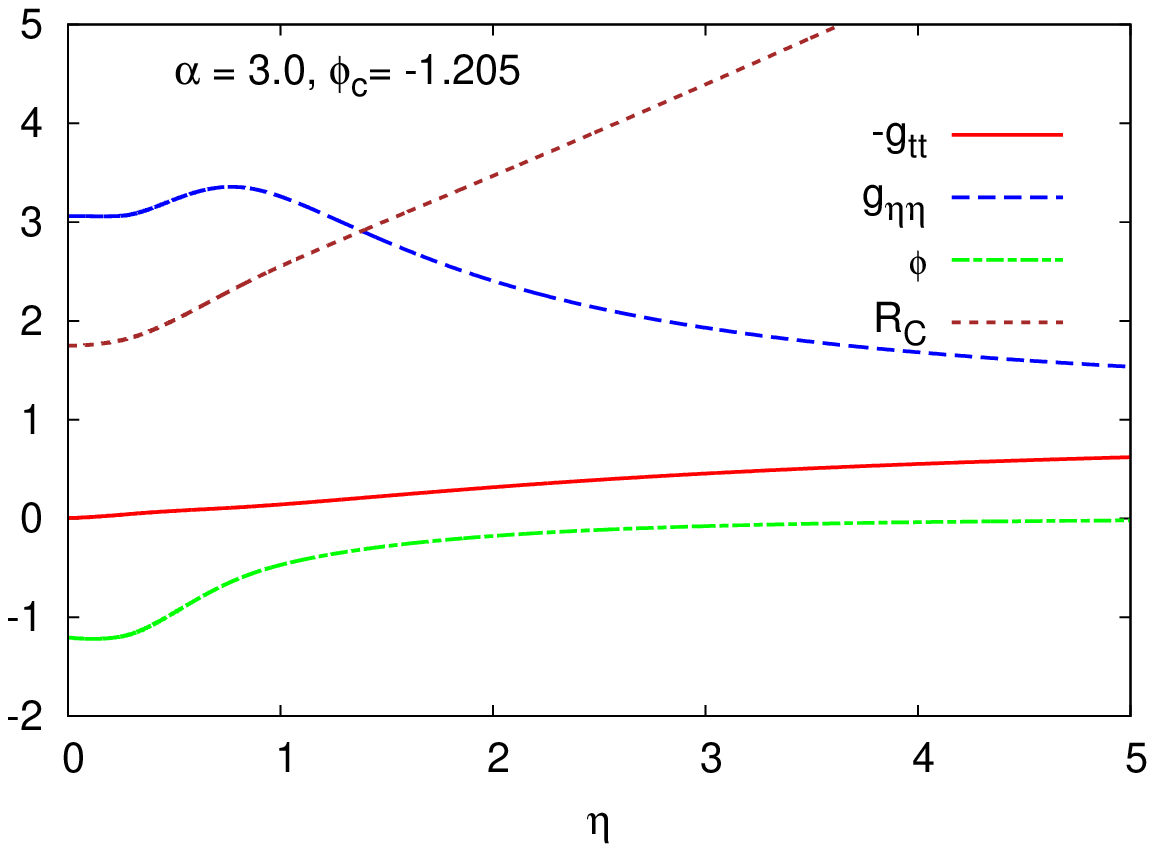}
\end{center}
\caption{
Metric components $-g_{tt}$ and $g_{\eta\eta}$, scalar field $\phi$,
and circumferential radial coordinate $R_C$ vs radial coordinate $\eta$
for a set of self-interacting wormhole solutions with parameters $(\alpha,\phi_c)$:
(a) $(0.8,-0.4)$, (b) $(1.1,-0.4)$, (c) $(1.4,-0.067)$, (d) $(1.4,-0.95)$, 
(e) $(1.7,-1.02)$, (f) $(3.0,-1.205)$.
}
\label{fig_solutions}
\end{figure}

We exhibit in Fig.~\ref{fig_solutions} the profile functions for a set of wormholes solutions
with self-interaction potential.
Shown are the metric components $-g_{tt}$ and $g_{\eta\eta}$, the scalar field $\phi$,
and the circumferential radial coordinate $R_C$ vs the radial coordinate $\eta$.
For the two free parameters, the coupling constant $\alpha$
and the value of the scalar field $\phi_c$ at the center,
we have selected the values
(a) $(\alpha,\phi_c)=(0.8,-0.4)$, a value for $\alpha$ close to its minimal value;
(b) $(1.1,-0.4)$, a larger value of $\alpha$;
(c) $(1.4,-0.067)$, a value of $\phi_c$ close to the black hole limit;
(d) $(1.4,-0.95)$, a small value of $\phi_c$;
(e) $(1.7,-1.02)$, another small value of $\phi_c$;
(f) $(3.0,-1.205)$,  a value for $\alpha$ close to its maximal value.

The metric function $-g_{tt}$ is always monotonically rising from a small value
at the center to its asymptotic value.
The metric function $g_{\eta\eta}$ is typically not monotonic, but exhibits a minimum
at the center, from where it rises to its maximum, before approaching its asymptotic value.
When the black hole limit is approached, however, it becomes monotonic
with its maximum at the center approaching the black hole value of  $g_{\eta\eta}(0)=4$
(see Fig.~\ref{fig_solutions}(c)).
The scalar field becomes very small as the black hole limit is approached.
But is never assumes large absolute values. These are on the order of one or smaller.
Of interest is also the circumferential coordinate $R_C$. 
As the black hole limit is approached, the value of $R_C$ at the center approaches
$R_C(0)=2$ (see Fig.~\ref{fig_solutions}(c)).
While $R_C$ is monotonically increasing
in the case of wormhole solutions with a single throat, it has a maximum at the
center for wormhole solutions with an equator and a double throat.

\subsection{Domain of existence}

\begin{figure}[h!]
\begin{center}
(a)\includegraphics[width=.47\textwidth, angle =0]{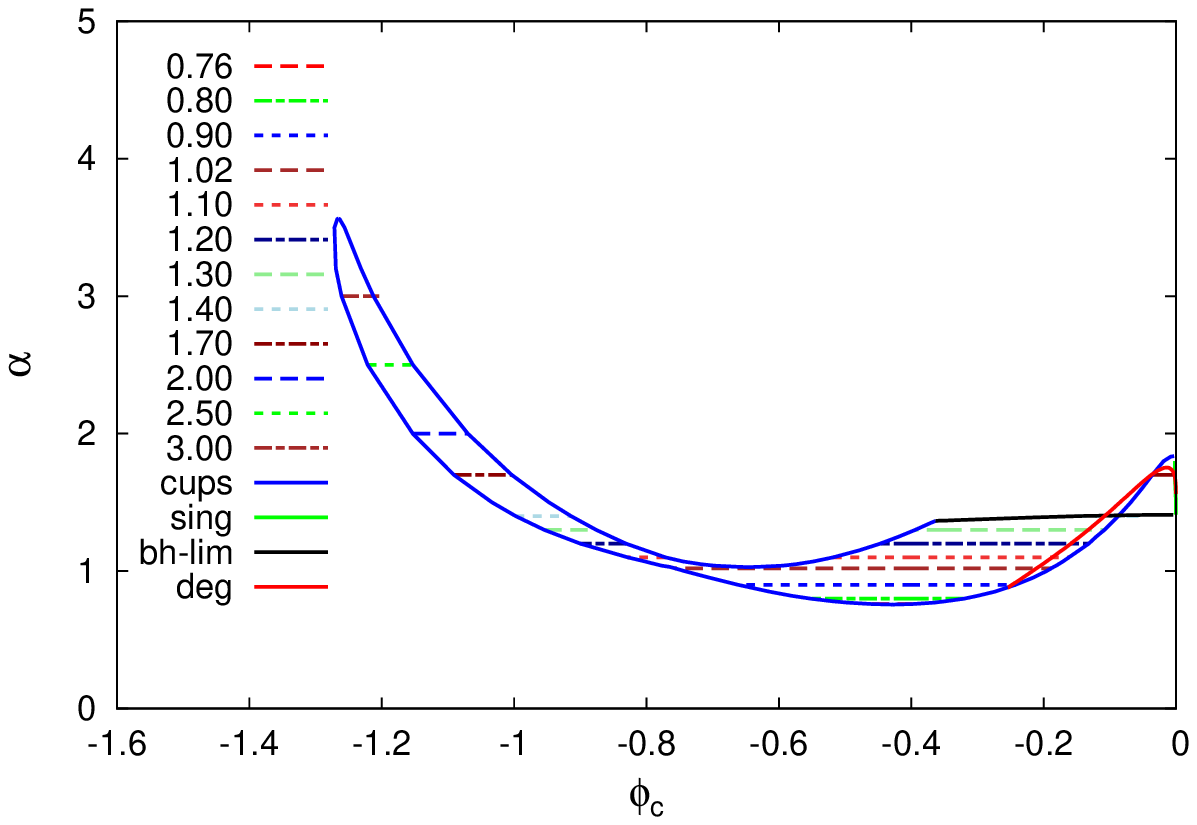}
(b)\includegraphics[width=.47\textwidth, angle =0]{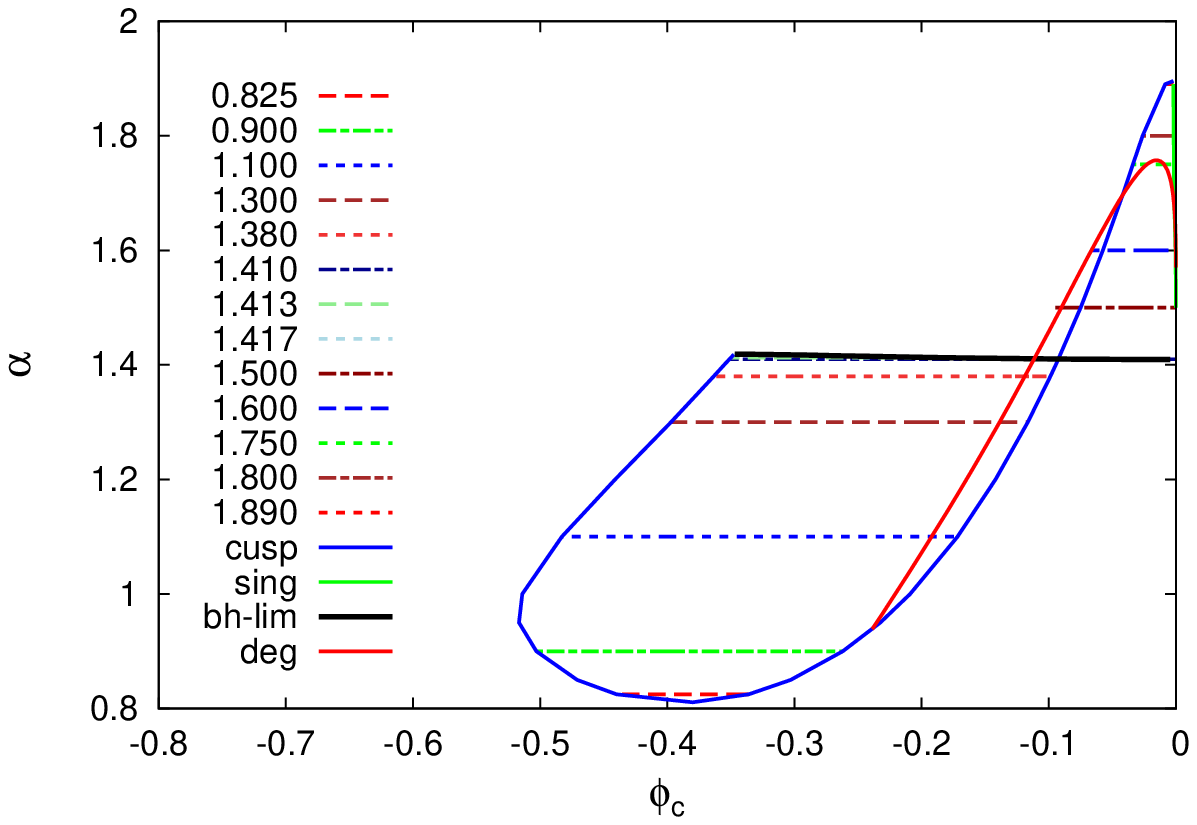}
\\
(c)\includegraphics[width=.47\textwidth, angle =0]{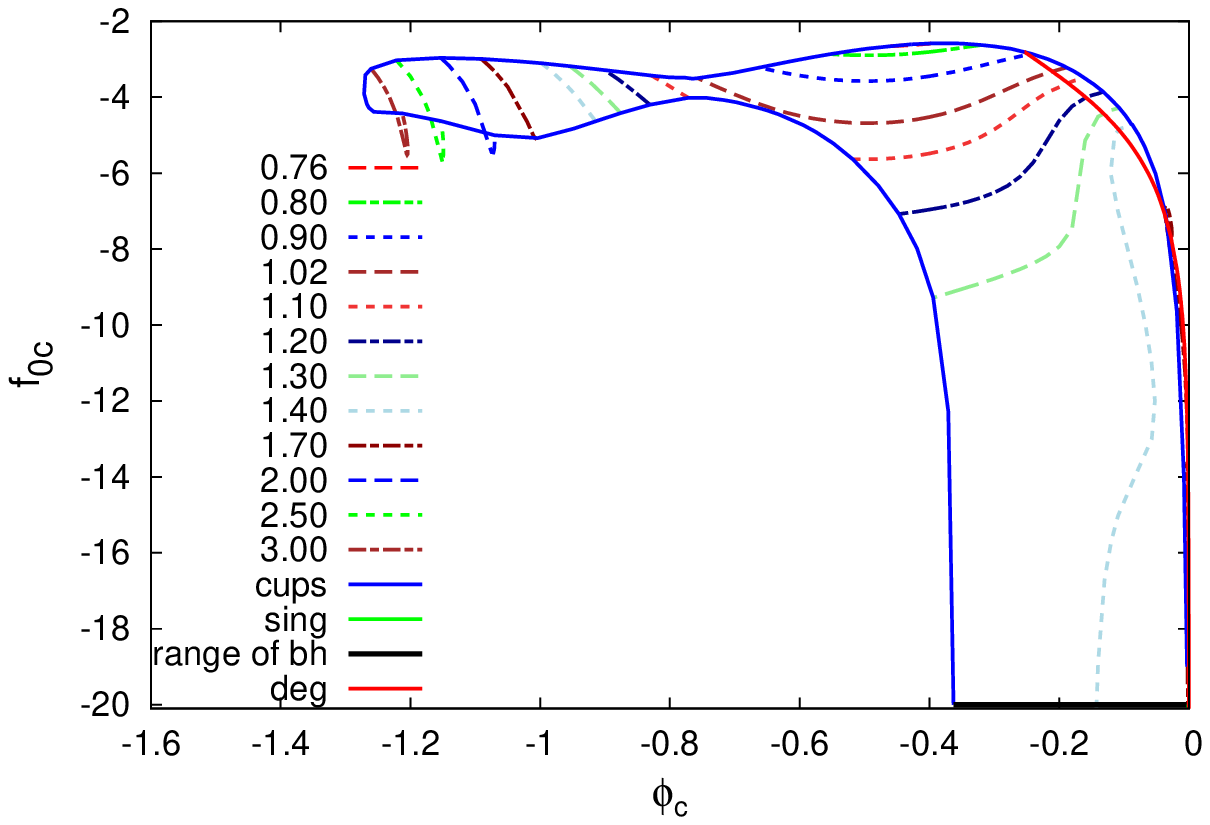}
(d)\includegraphics[width=.47\textwidth, angle =0]{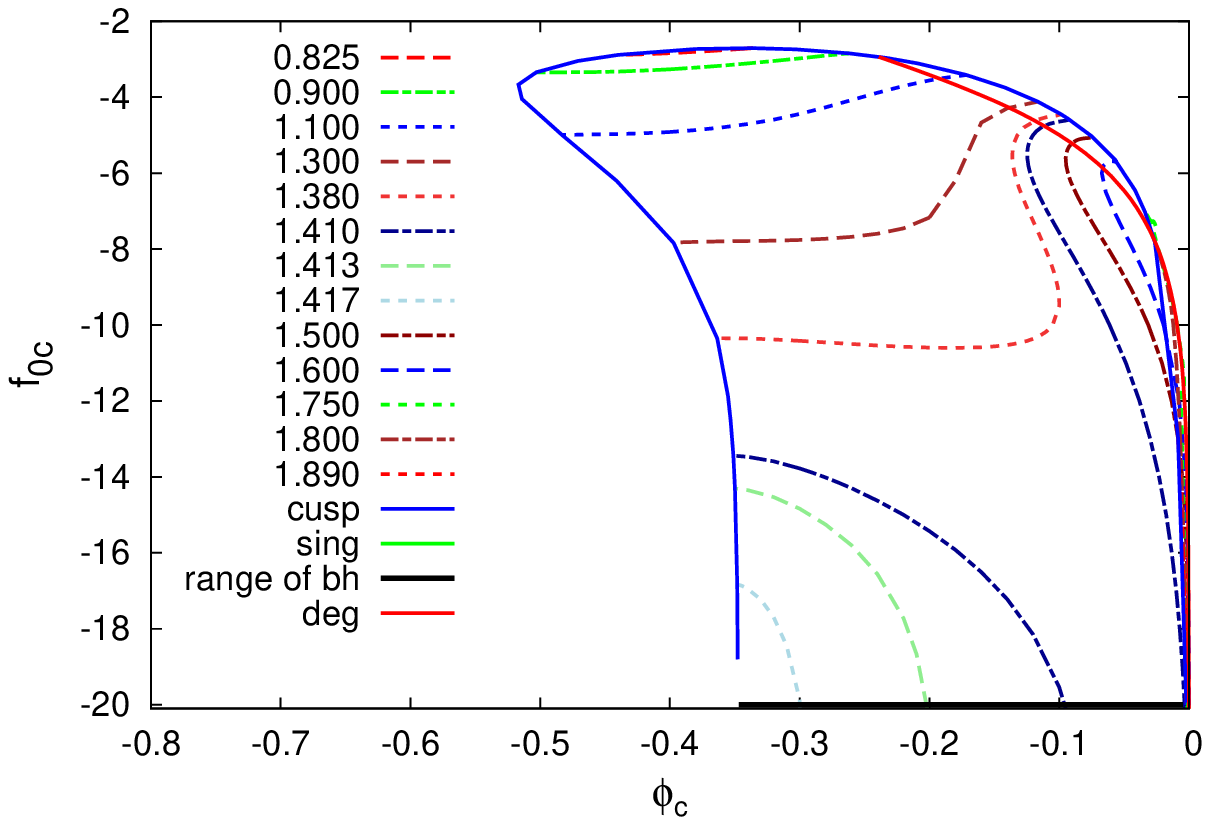}
\\
(e)\includegraphics[width=.47\textwidth, angle =0]{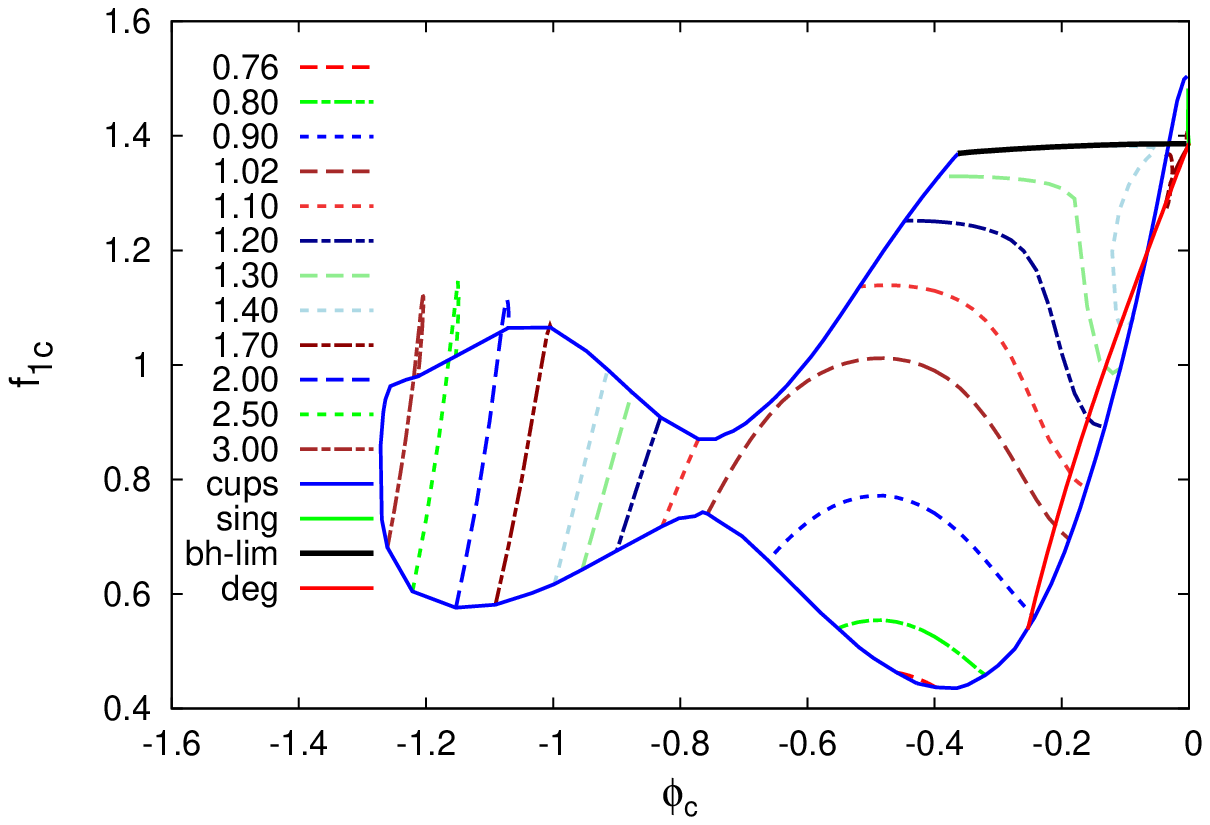}
(f)\includegraphics[width=.47\textwidth, angle =0]{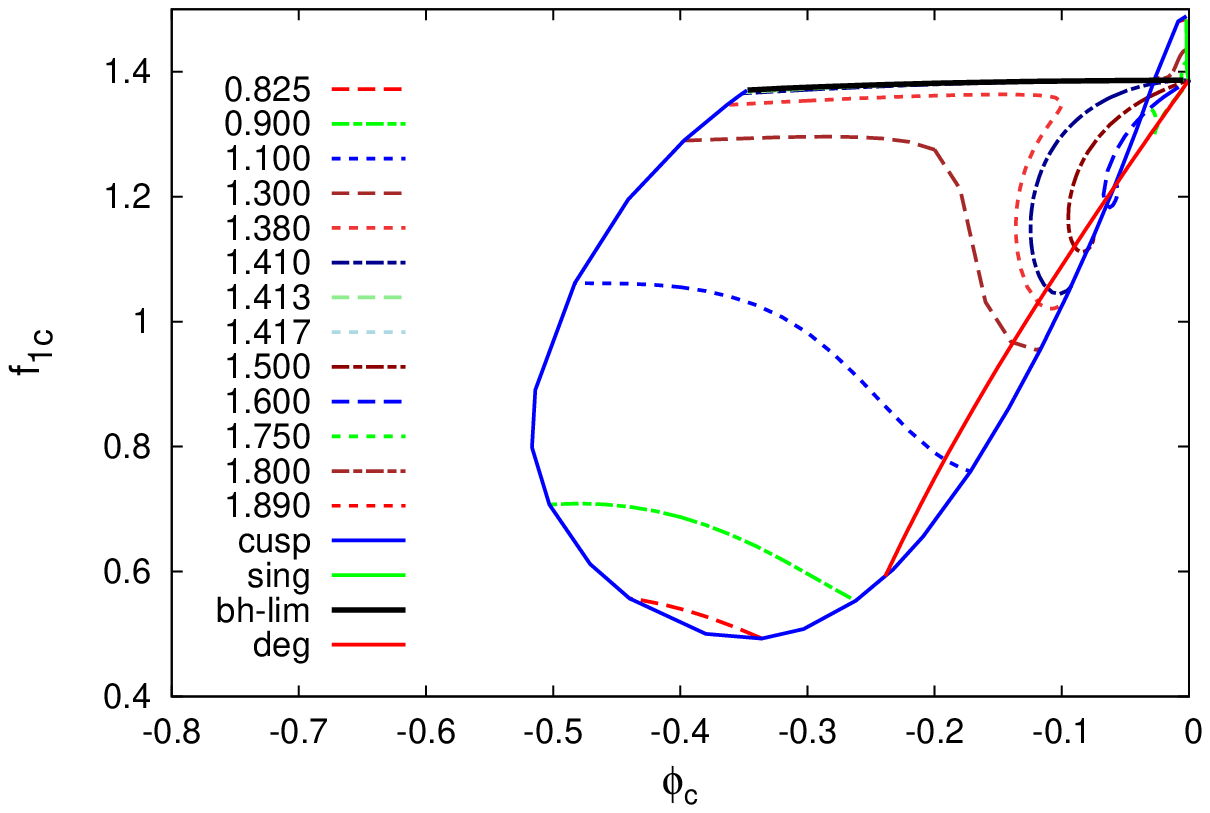}
\end{center}
\caption{
Domain of existence of wormhole solutions for a set of fixed values of the
coupling constant $\alpha$. Various quantities are shown vs the value
of the scalar field at the center $\phi_c$:
(a) and (b) coupling constant $\alpha$,  
(c) and (d) value of the metric function $f_0$ at the center, $f_{0c}$, 
(e) and (f) value of the metric function $f_1$ at the center, $f_{1c}$
[left column: with self-interaction, right column: with mass term only].
Also shown are the limiting solutions, scalarized EsGB black holes (black), 
singular solutions (green), cusp singularities (blue),
and the degenerate wormhole solutions (red).
}
\label{fig_domain1}
\end{figure}
\begin{figure}[h!]
\begin{center}
(a)\includegraphics[width=.47\textwidth, angle =0]{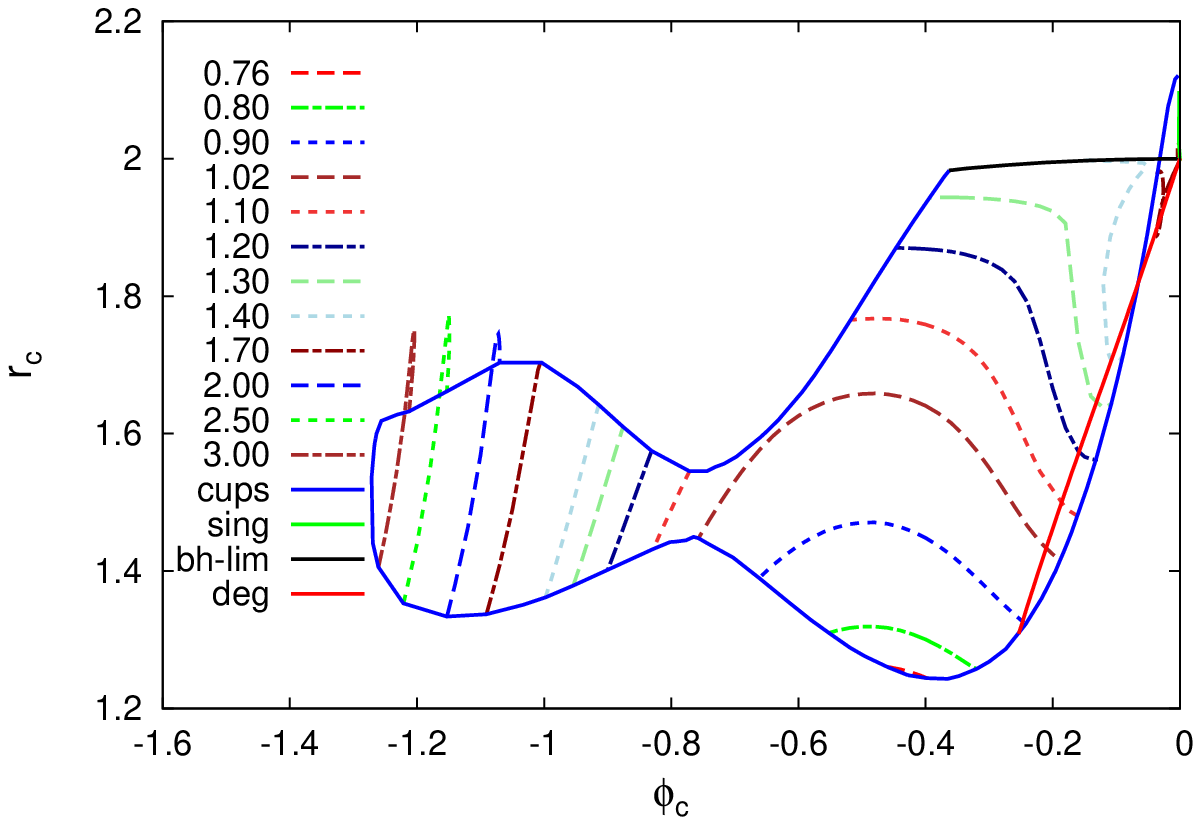}
(b)\includegraphics[width=.47\textwidth, angle =0]{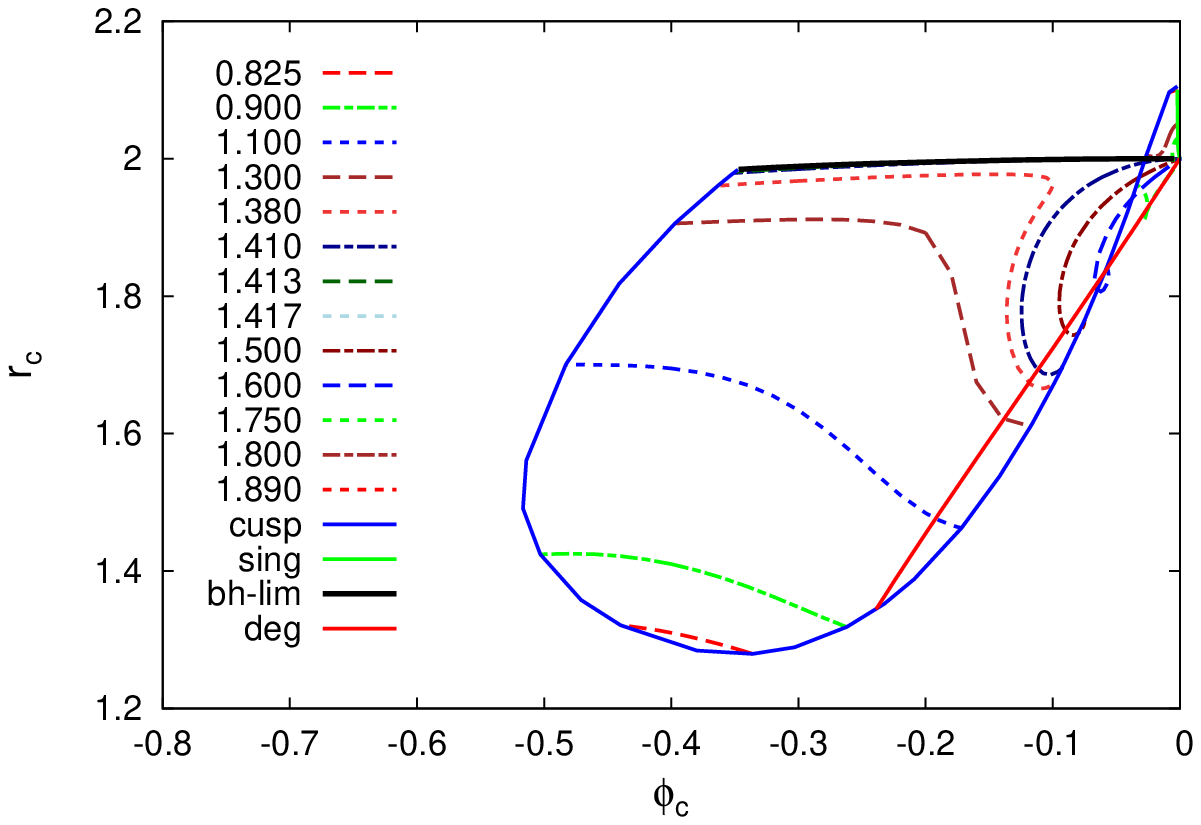}
\\
(c)\includegraphics[width=.47\textwidth, angle =0]{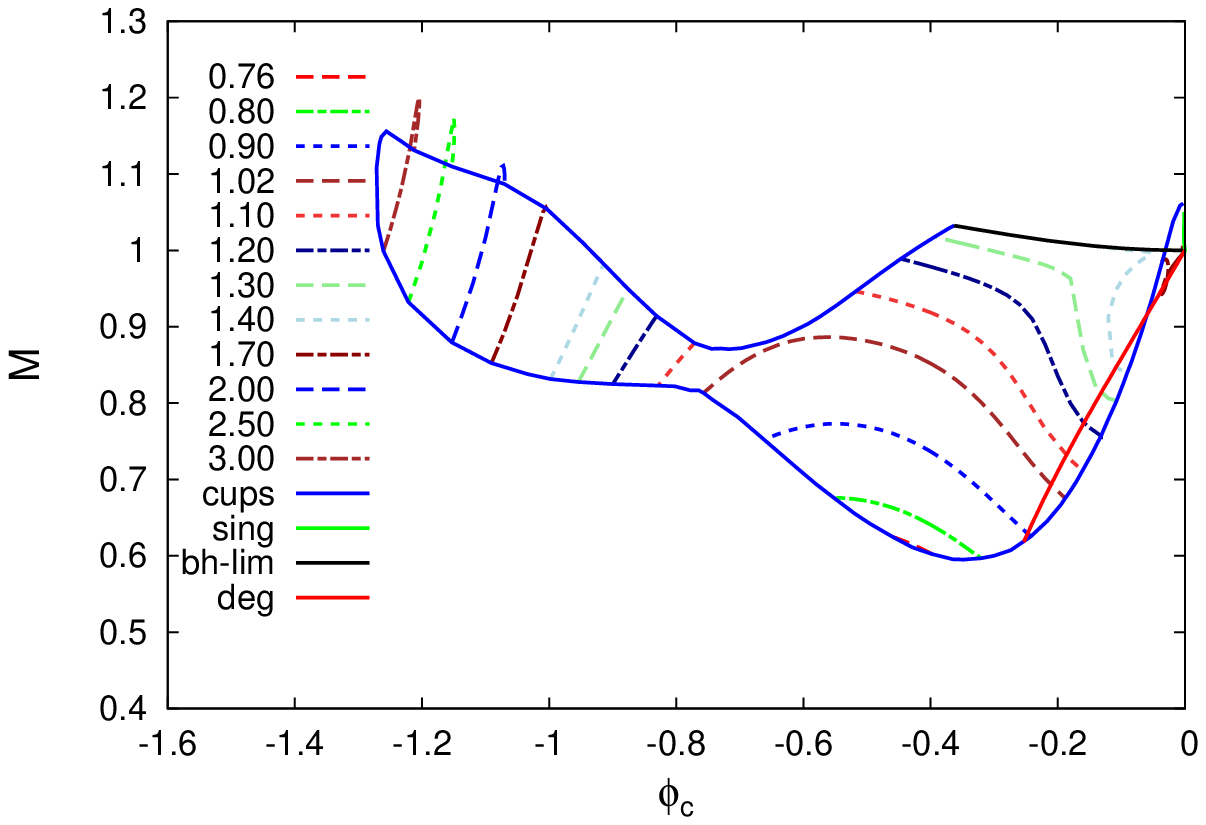}
(d)\includegraphics[width=.47\textwidth, angle =0]{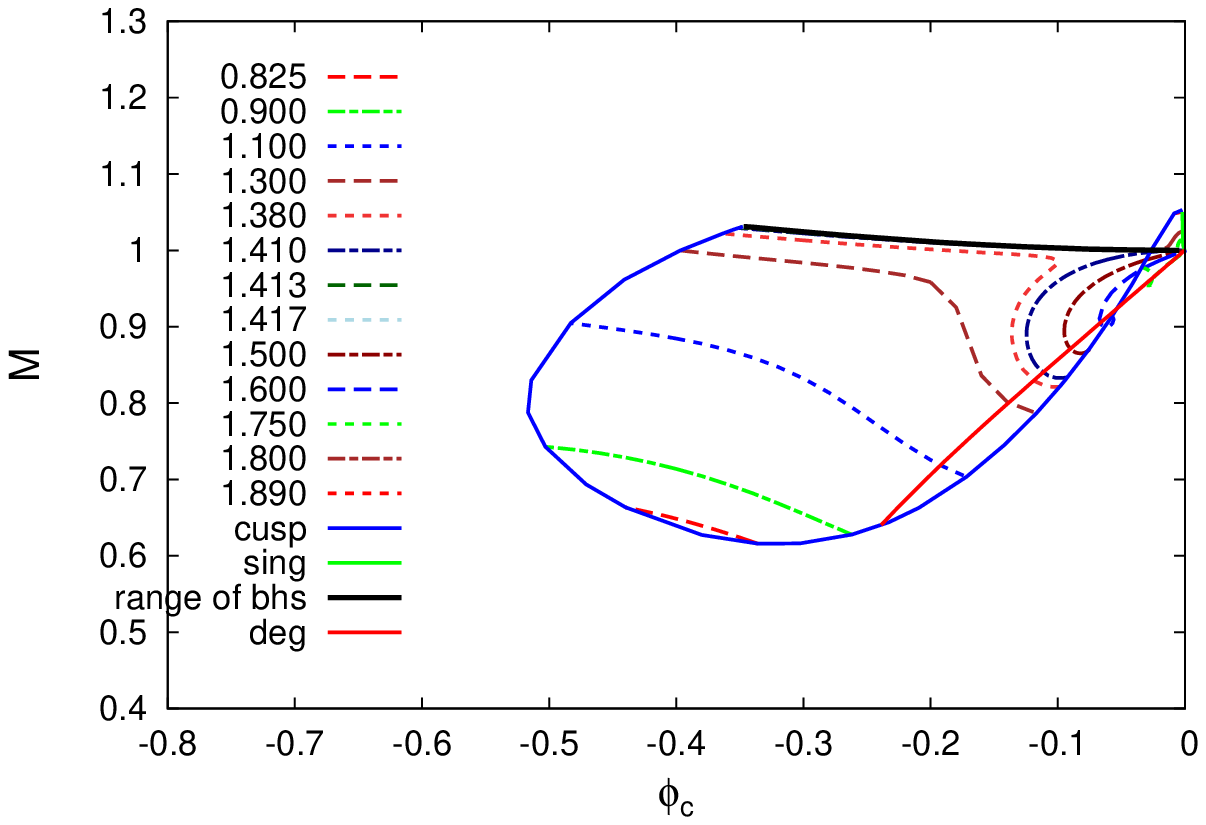}
\\
(e)\includegraphics[width=.47\textwidth, angle =0]{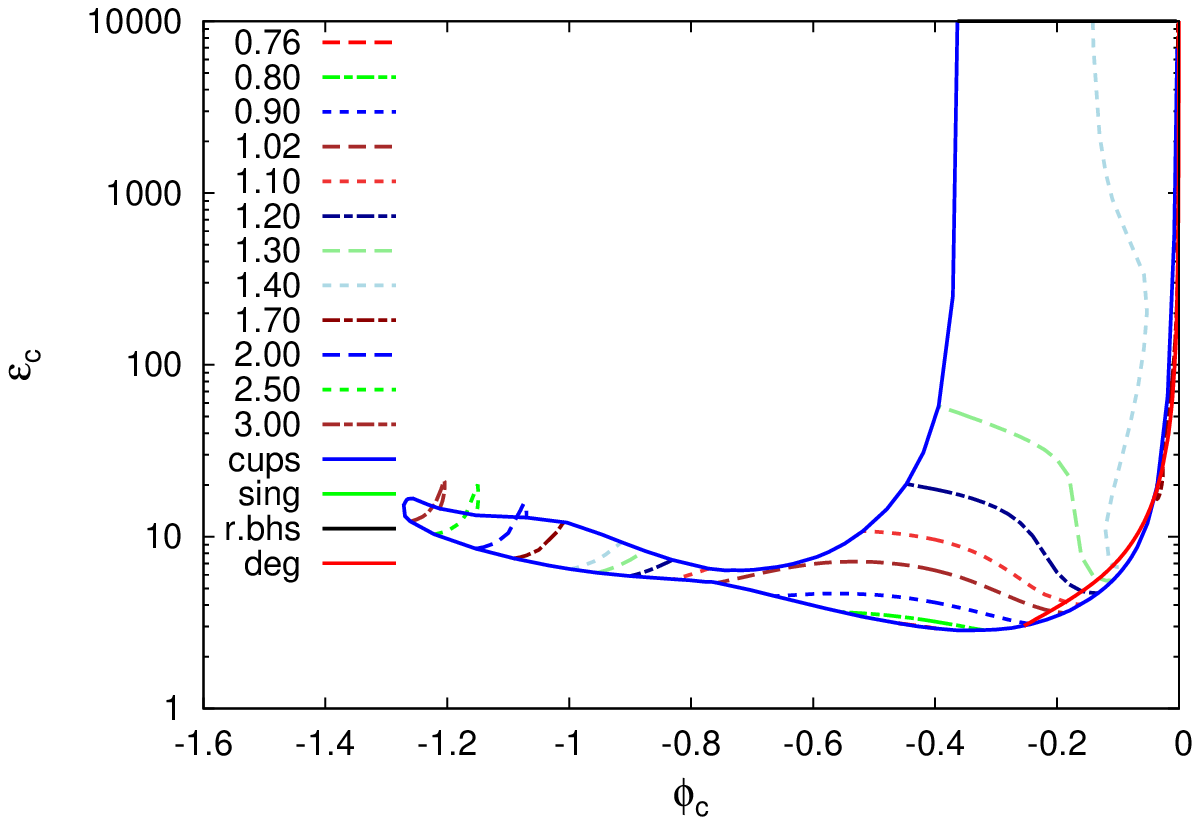}
(f)\includegraphics[width=.47\textwidth, angle =0]{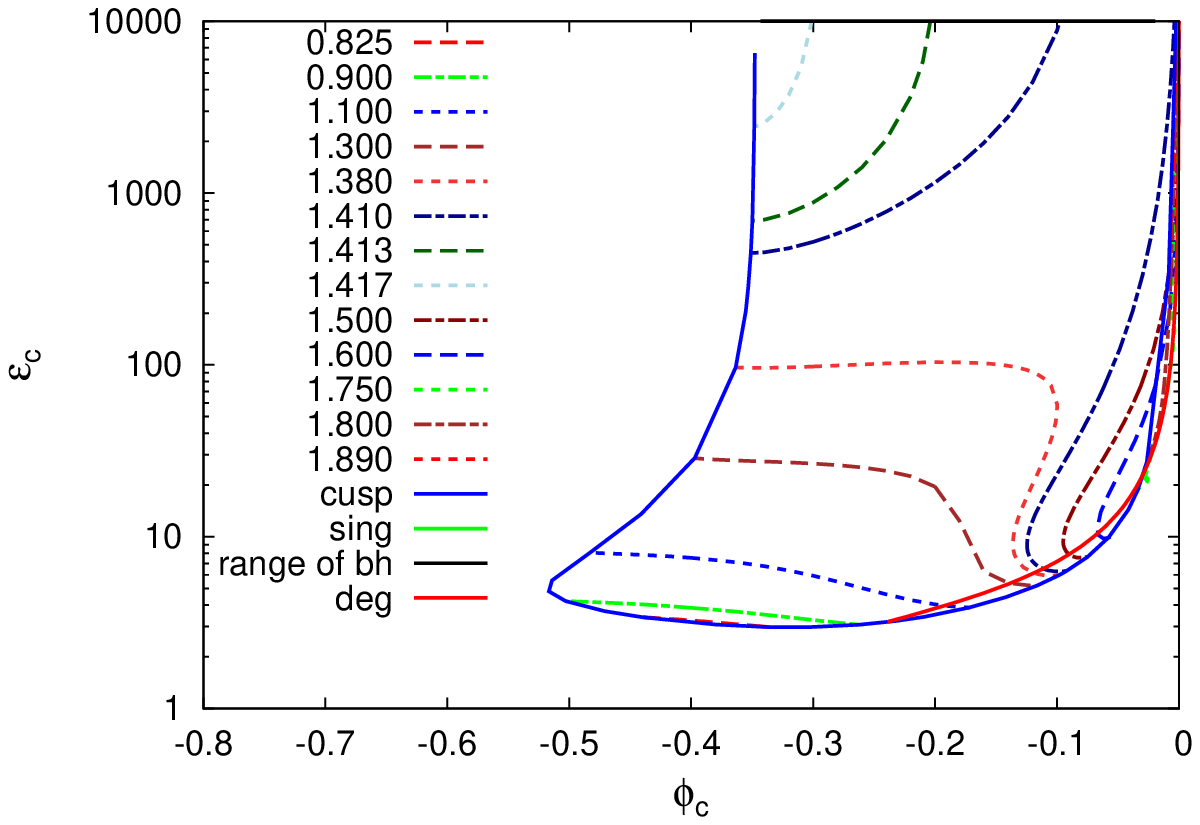}
\end{center}
\caption{
Domain of existence of wormhole solutions for a set of fixed values of the
coupling constant $\alpha$. Various quantities are shown vs the value
of the scalar field at the center $\phi_c$:
(a) and (b)  value of the circumferential radius $R_C$ at the center, $r_c$, 
(c) and (d) value of the mass $M$, 
(e) and (f) value of the energy density at the center, $\varepsilon_c$
[left column: with self-interaction, right column: with mass term only].
Also shown are the limiting solutions, scalarized EsGB black holes (black), 
singular solutions (green), cusp singularities (blue),
and the degenerate wormhole solutions (red).
}
\label{fig_domain2}
\end{figure}

Wormhole solutions exist for a limited range of parameter values. 
The domain of existence of wormhole solutions is illustrated in 
Fig.~\ref{fig_domain1} for wormholes with mass and self-interaction (left column)
and wormholes with a mass term only (right column).
As seen in Fig.~\ref{fig_domain1}(a), where the coupling constant $\alpha$
is shown versus the value of the scalar field at the center $\phi_c$,
wormhole solutions arise for a minimal value of the coupling constant $\alpha$,
$\alpha_{\text{min}} \approx 0.76$.
As $\alpha$ increases, the corresponding families of solutions expand,
until a critical value of $\alpha$, $\alpha_{\text{cr}} \approx 1.02$, is reached.

Further increase of $\alpha$ then leads to two disconnected branches of 
wormhole solutions.
The branches with the smaller values of $\phi_c$ can be continued
up to $\alpha_{\text{max}} \approx 3.56$.
The branches with the larger values of $\phi_c$, on the other hand,
cannot be continued that far. They end at a limiting value of
$\alpha_{\text{lim}} \approx 1.83$.
The domain of existence is delimited by solutions with cusp singularities (blue),
solutions with singularities at the center (green), and scalarized EsGB black hole solutions (black),
reaching all the way up to their bifurcation point from the Schwarzschild black hole. 
The limiting solutions will be discussed further below.

For wormholes with a mass term only, the coupling constant $\alpha$
is shown versus the value of the scalar field at the center $\phi_c$
in Fig.~\ref{fig_domain1}(b).
We immediately note, that the domain of existence basically only consists
of one part of the domain of existence of wormholes with self-interaction.
This part more or less agrees with the right hand side of the 
domain shown in Fig.~\ref{fig_domain1}(a), and thus the larger values of $\phi_c$.
Thus the effect of the self-interaction is to allow for 
wormhole solutions with much smaller values of $\phi_c$
and much larger values of the coupling constant $\alpha$.

We exhibit
the value of the metric function $f_0$ at the center, $f_{0c}$, 
in Fig.~\ref{fig_domain1}(c) and (d), and 
the value of the metric function $f_1$ at the center, $f_{1c}$,
in Fig.~\ref{fig_domain1}(e) and (f),
with (c) and (e) ((d) and (f)) representing wormholes with (without) self-interaction.
Again, we see, how the presence of the self-interaction leads to an opening of the
left hand boundary of the wormhole solutions with mass tern only at its leftmost point
to allow a new region of solutions to be present,
where considerably smaller values of  $f_{0c}$ and  $f_{1c}$ are reached
and much higher values of the coupling constant.

In particular, we note, that in the left upper region,
i.e., in the small $\phi_c$ region
present only for wormholes with self-interaction,
there are two branches of $f_{0c}$ and  $f_{1c}$
for a given value of $\alpha$.
The first branch starts from a cusp singularity while the second
branch ends in a cusp singularity.
In the large $\phi_c$ region, on the other hand, the value $f_{0c}$
decreases without bound as the black hole limit
and thus a horizon is approached.
In contrast, the value $f_{1c}$ tends towards a finite limiting black hole value, 
since we are not employing Schwarzschild-like coordinates.
In the Schwarzschild limit it approaches in our isotropic coordinates the value $f_{1c}=2 \ln 2$.

This is consistent with the limiting values of the
circumferential radius at the center, which approaches 
in the Schwarzschild limit $r_c=2$.
This is seen in Fig.~\ref{fig_domain2},
where the value of the circumferential radius $R_C$ at the center, $r_c$,
is shown with (a) and without (b) self-interaction,
versus the value of the scalar field at the center $\phi_c$.
We note that the leftmost part of the domain of existence,
present due to the self-interaction, does not feature
larger or smaller circumferential throat radii
than those already present without self-interaction.

The chosen scaling corresponds to a mass of the 
limiting Schwarzschild black hole of $M=1$.
This is seen in Fig.~\ref{fig_domain2}(c) and (d),
where the mass $M$ of the wormhole solutions is shown versus
the value of the scalar field at the center $\phi_c$.
We note, that the overall variation of the mass 
is moderate.
In the region due to self-interaction, however,
the mass can get bigger than in the remaining domain
of existence.
Thus the presence of the self-interaction not only
increases the domain of existence but also
allows for larger values of the mass.
This is, in fact, analogous to the effect of the self-interaction
for boson stars \cite{Friedberg:1986tq,Kleihaus:2005me}.

Finally, in Fig.~\ref{fig_domain2}(e) and (f) we exhibit the 
matter energy density $\varepsilon$ in the thin shell at the center,
Eq.~(\ref{rhomat1}), assuming dust and choosing
$\lambda_1=\lambda_2$. Already for this simple choice
there is quite a large range of wormhole solutions,
where we find positive values for $\varepsilon$.
The additional set of wormholes due to the self-interaction
have all positive values of the matter energy density $\varepsilon$.
Thus these wormholes are constituted by physically allowed matter,
that is respecting the energy conditions.

In Figs.~\ref{fig_domain1} and \ref{fig_domain2} we also exhibit the line (red),
where the throat becomes degenerate, i.e., where the minimum of $R_C$
turns into a saddle point. 
The wormholes to the left of this line possess a single throat 
located at the center.
The wormholes to the right of this line possess an equator
at the center that is surrounded by a throat on each side.
The geometry of these solutions is discussed further below.
This region with wormholes with equators is basically
unchanged by the presence of the self-interaction.



\subsection{Limits}

\begin{figure}[t!]
\begin{center}
(a)\includegraphics[width=.47\textwidth, angle =0]{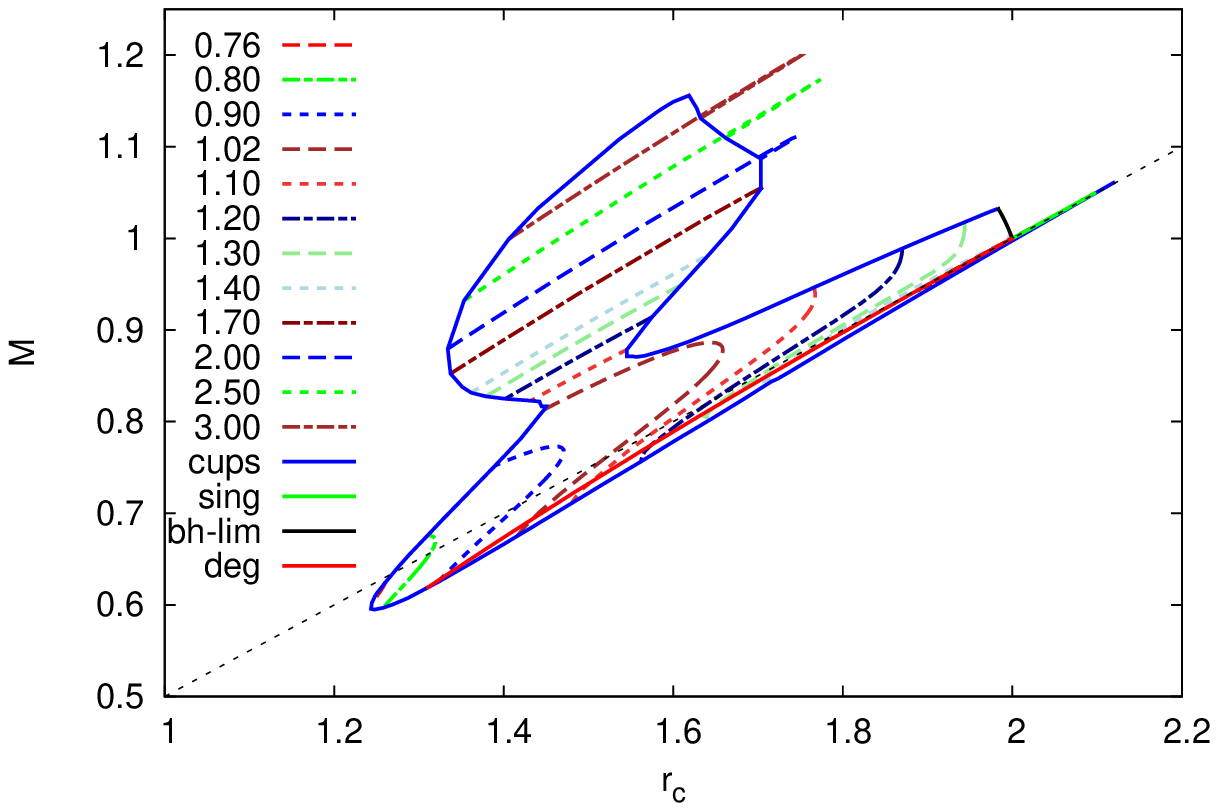}
(b)\includegraphics[width=.47\textwidth, angle =0]{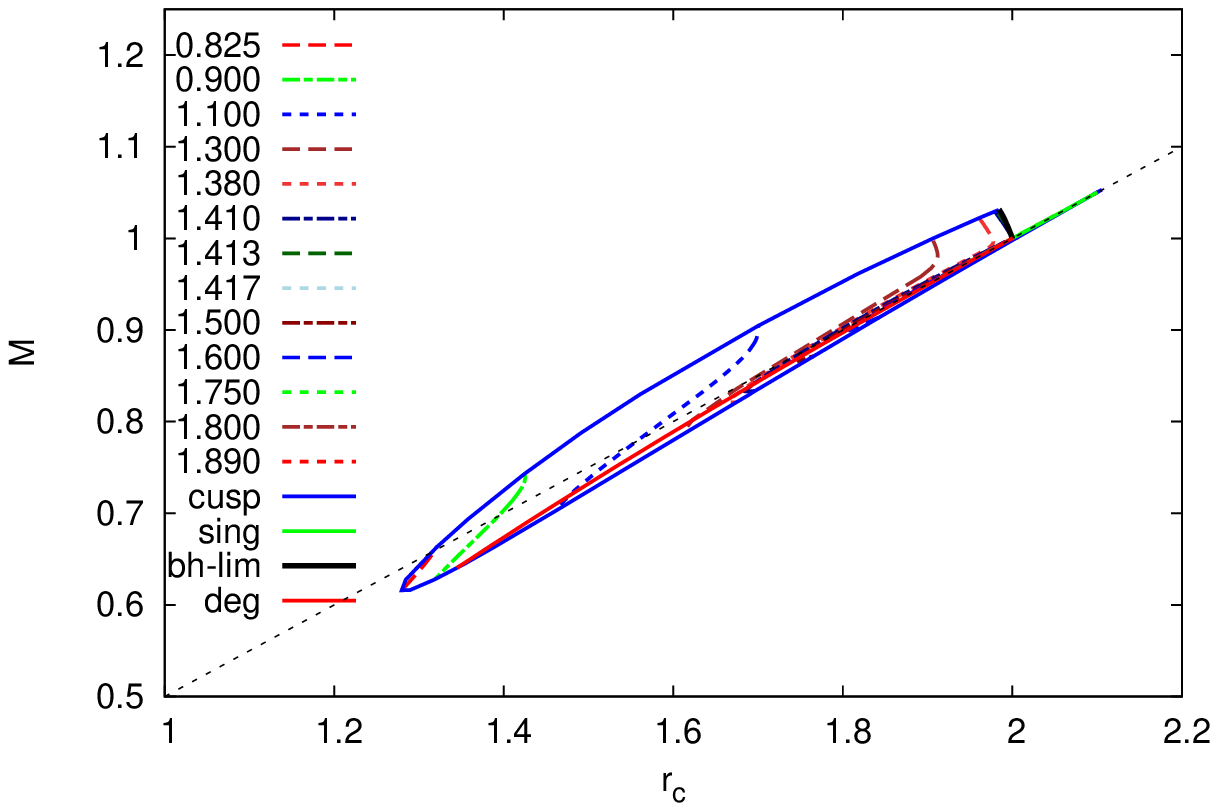}
(c)\includegraphics[width=.47\textwidth, angle =0]{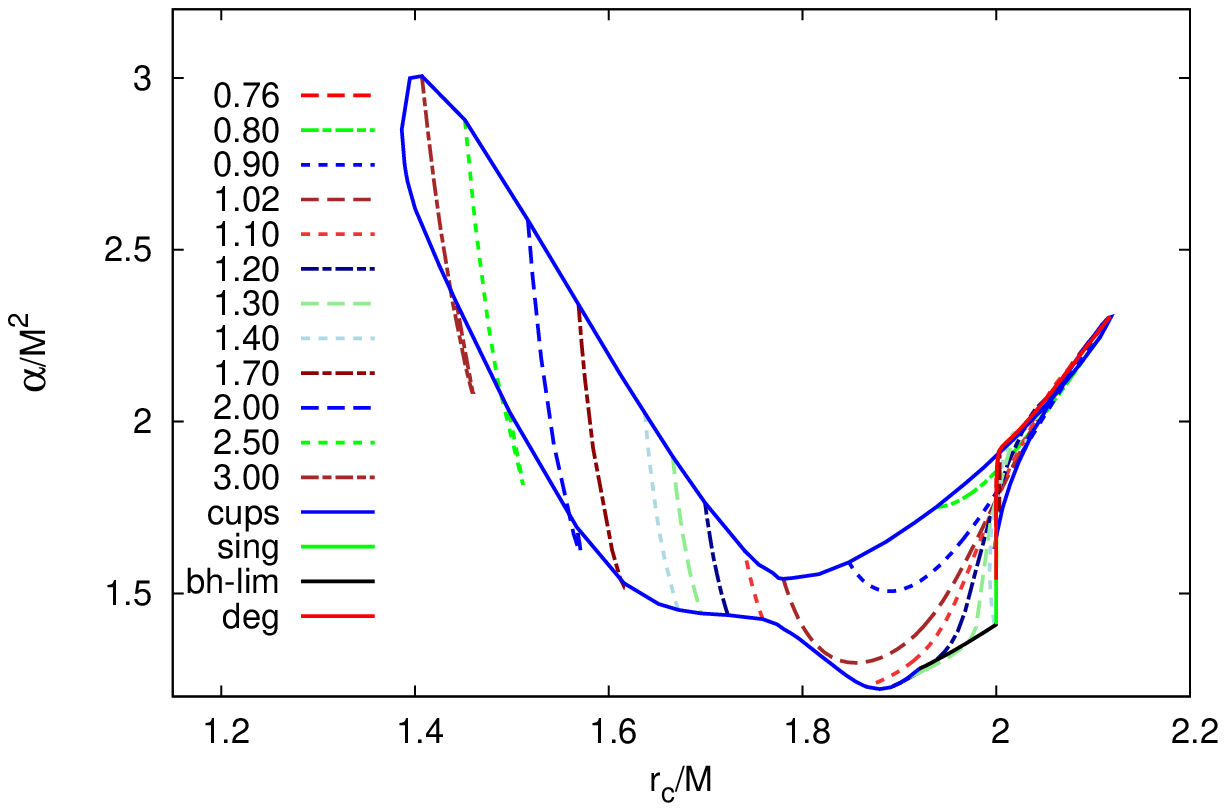}
(d)\includegraphics[width=.47\textwidth, angle =0]{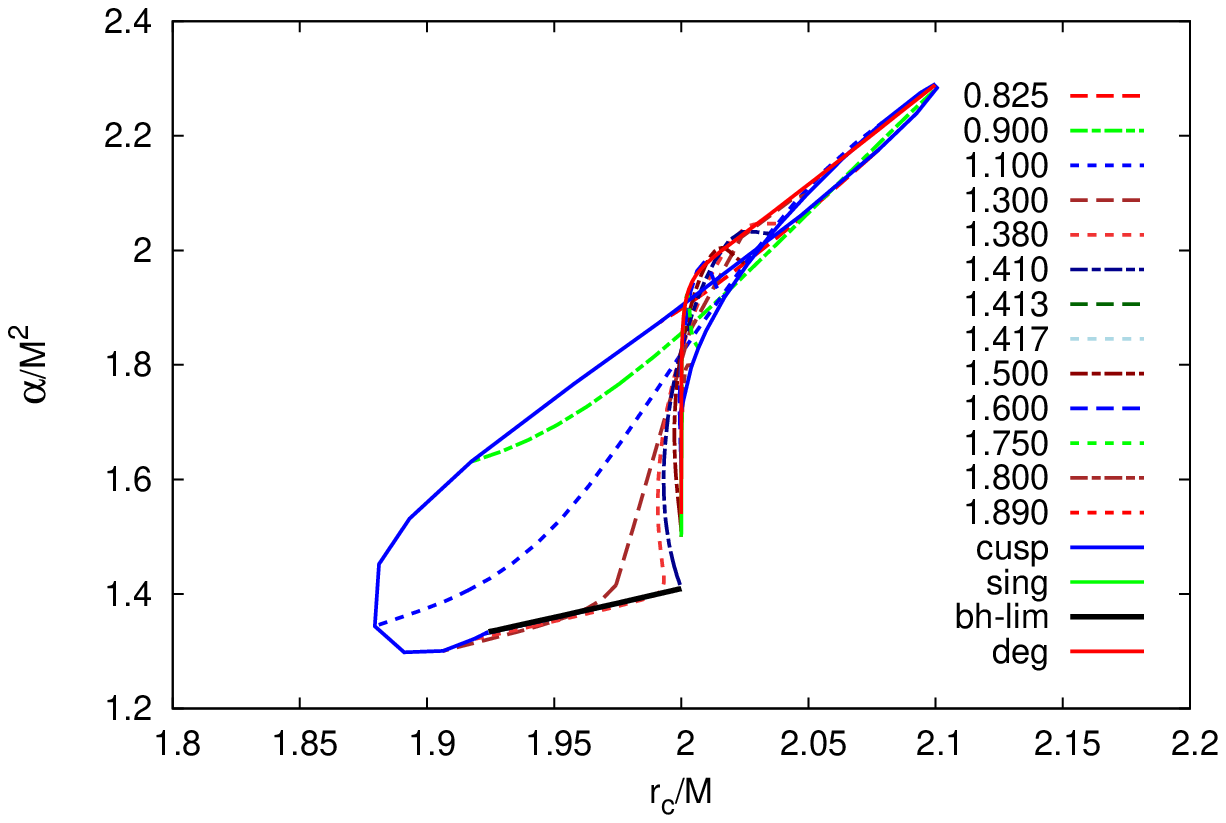}
\end{center}
\caption{
(a) and (b) Mass $M$ vs circumferentical radius $r_c$ at the center
for fixed values of the coupling constant $\alpha$;
(c) and (d) scaled coupling constant $\alpha/M^2$ vs scaled
circumferentical radius $r_c/M$ at the center
for the same set of $\alpha$ values
[left column: with self-interaction, right column: with mass term only].
Also shown are the limiting solutions, scalarized EsGB black holes (solid black)
[Schwarzschild black holes (dotted black)], 
singular solutions (green), cusp singularities (blue),
and the degenerate wormhole solutions (red).
}
\label{fig_scaled}
\end{figure}
\begin{figure}[h!]
\begin{center}
(a)\includegraphics[width=.47\textwidth, angle =0]{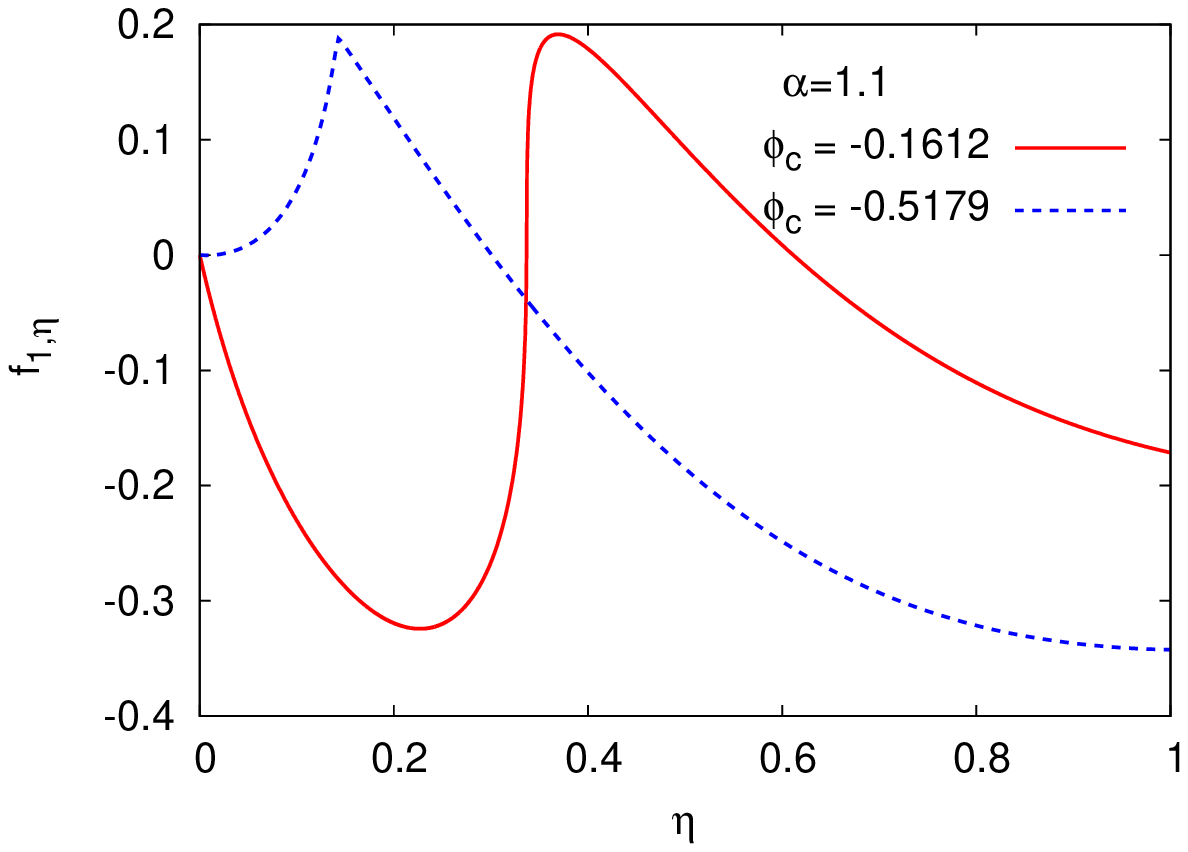}
(b)\includegraphics[width=.47\textwidth, angle =0]{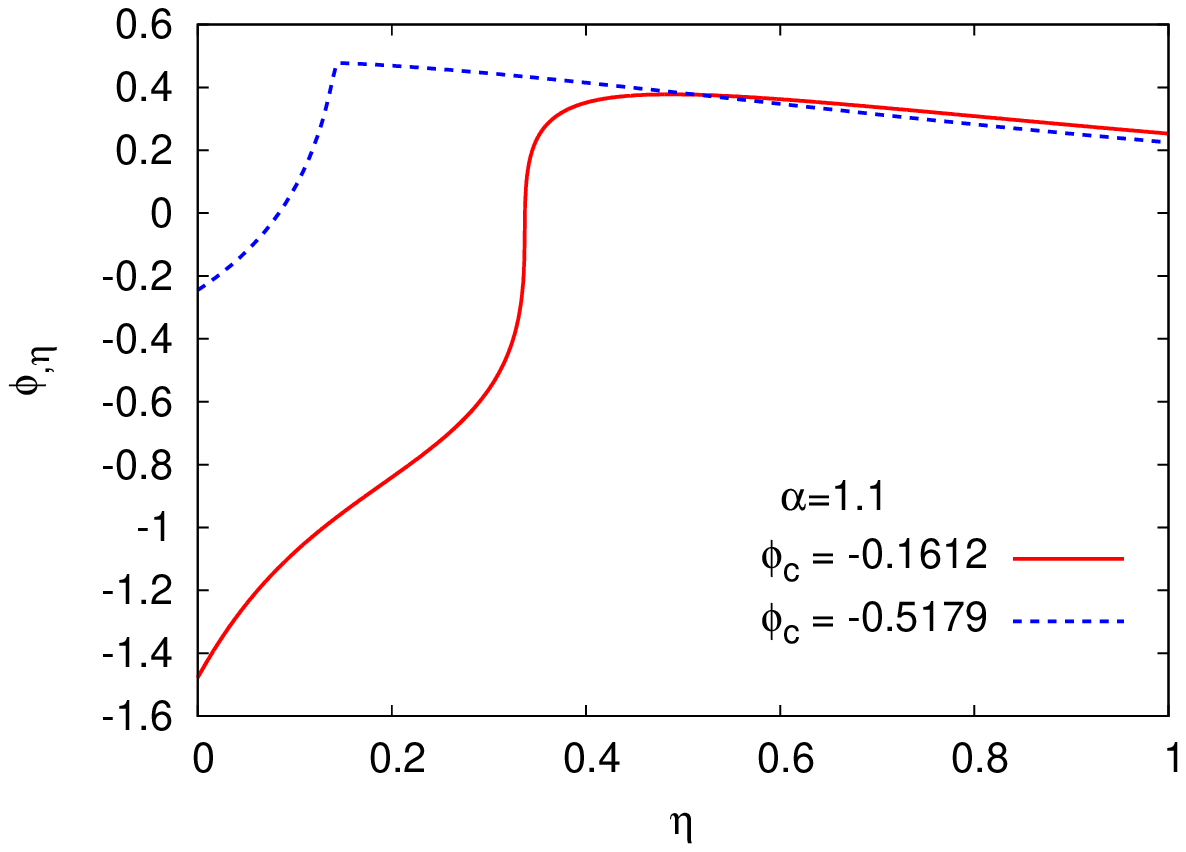}
\\
(c)\includegraphics[width=.47\textwidth, angle =0]{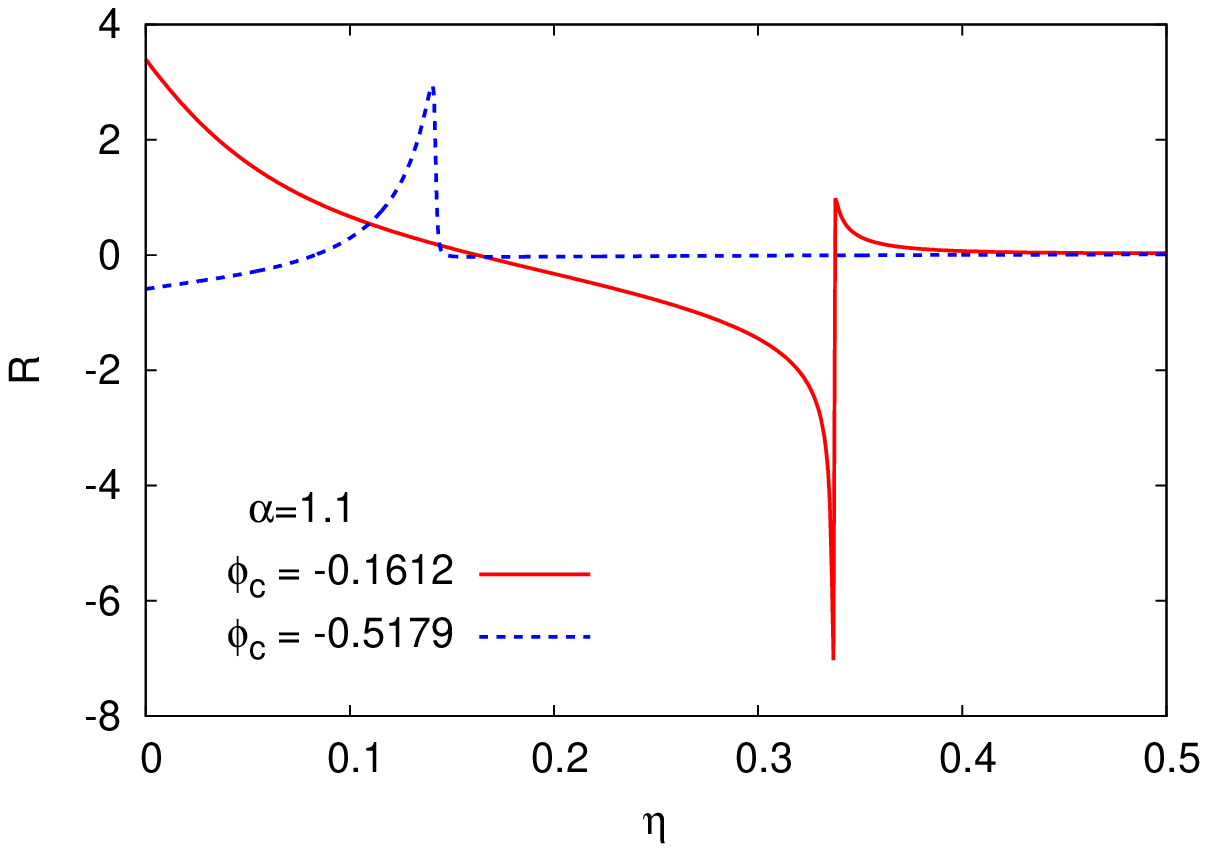}
(d)\includegraphics[width=.47\textwidth, angle =0]{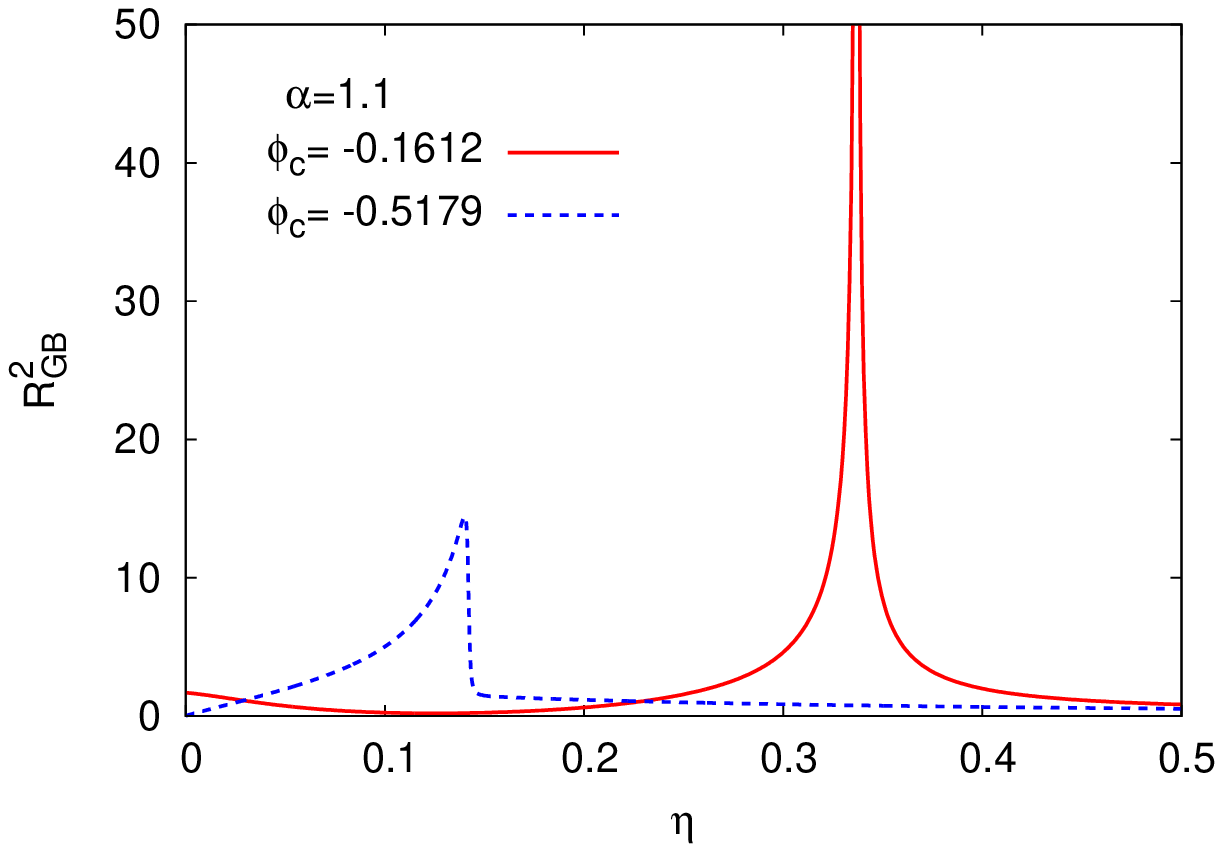}
\end{center}
\caption{Boundary of the domain of existence
(examples with self-interaction):
emergence of a cusp singularity at some value $\eta_\star$:
the first derivative of the metric function $f_1$ (a) 
and of the scalar field function $\phi$ (b)
for $\alpha=1.1$ and two values
of the scalar field at the center $\phi_c$: 
$f_{1,\eta\eta}$ and  $\phi_{,\eta\eta}$ develop a jump at some $\eta_\star$ (blue, $\phi_c=-0.5179$),
$f_{1,\eta\eta}$ and  $\phi_{,\eta\eta}$ diverge at some $\eta_\star$ (red, $\phi_c=-0.1612$);
scalar curvature $R$ (c) and Gauss-Bonnet term $R^2_{GB}$ (d)
vs radial coordinate $\eta$  for the same solutions.
}
\label{fig_limits}
\end{figure}
\begin{figure}[h!]
\begin{center}
(a)\includegraphics[width=.47\textwidth, angle =0]{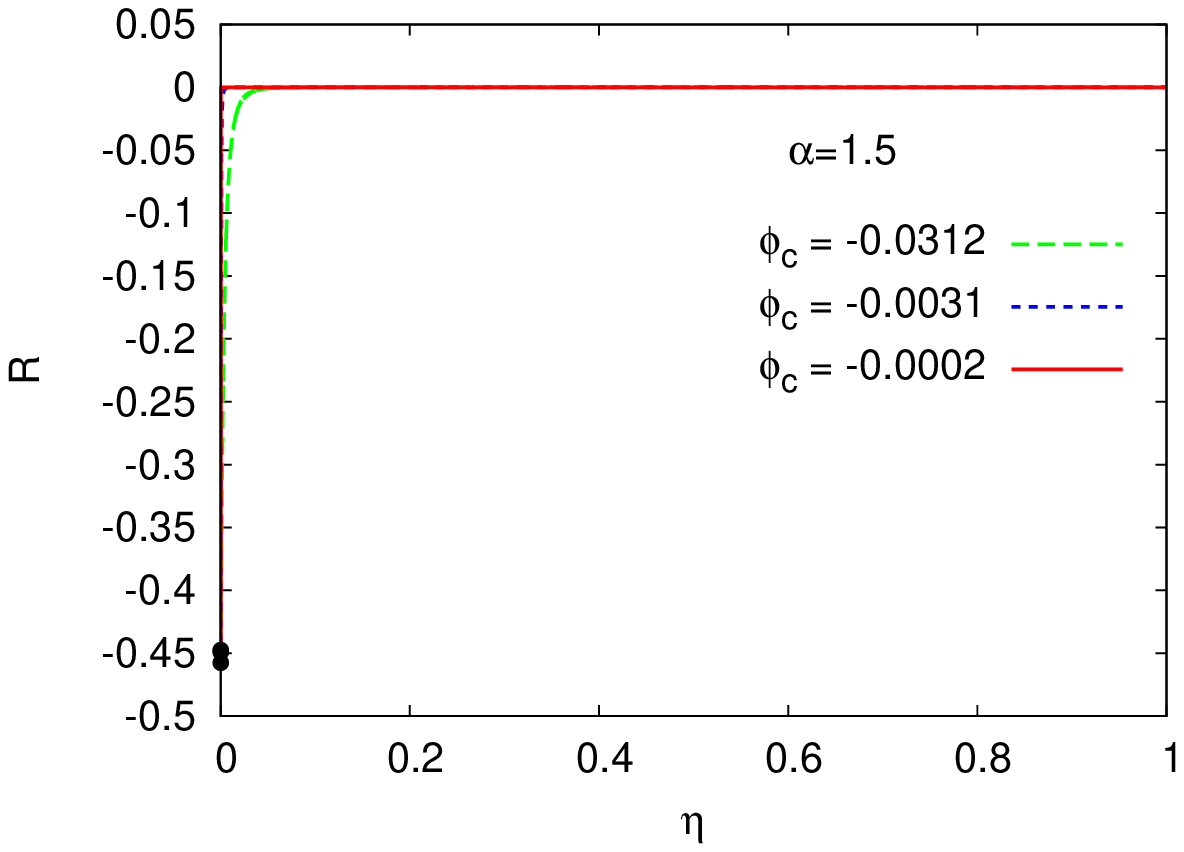}
(b)\includegraphics[width=.47\textwidth, angle =0]{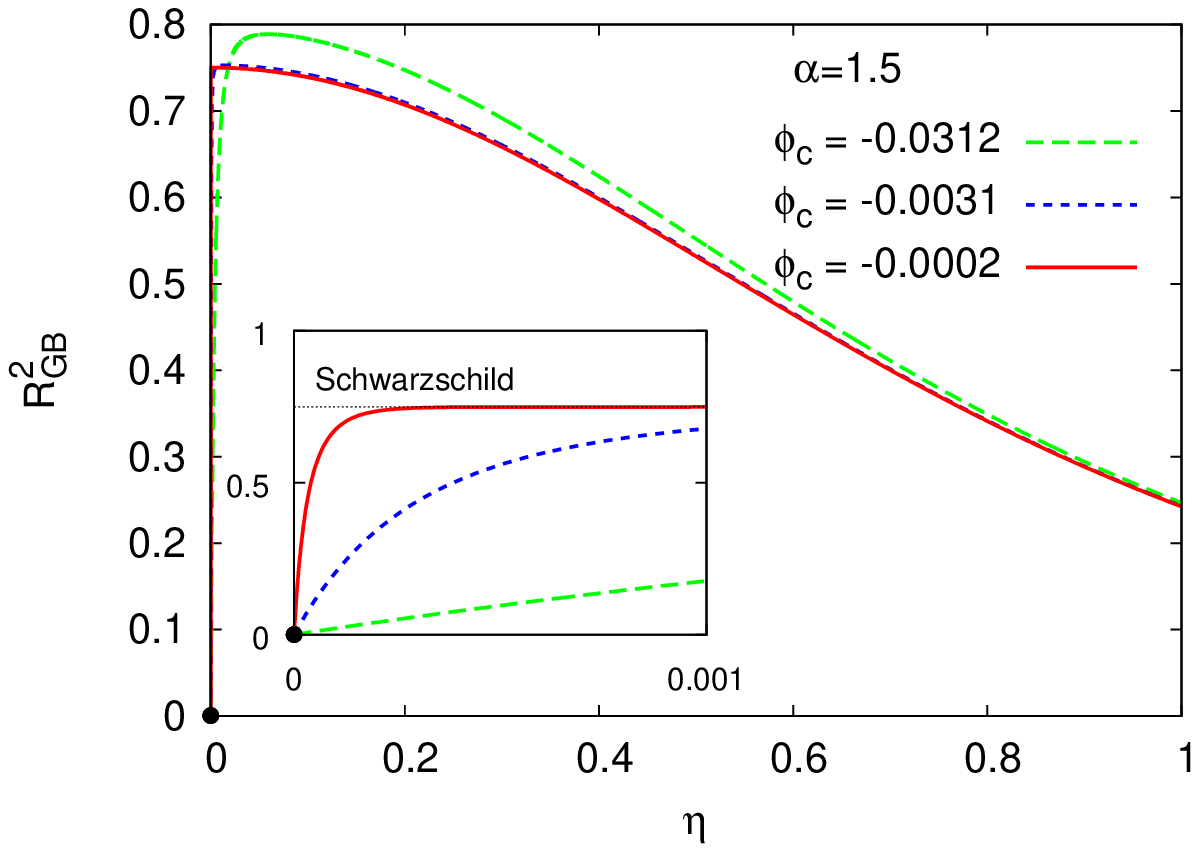}
\\
(c)\includegraphics[width=.47\textwidth, angle =0]{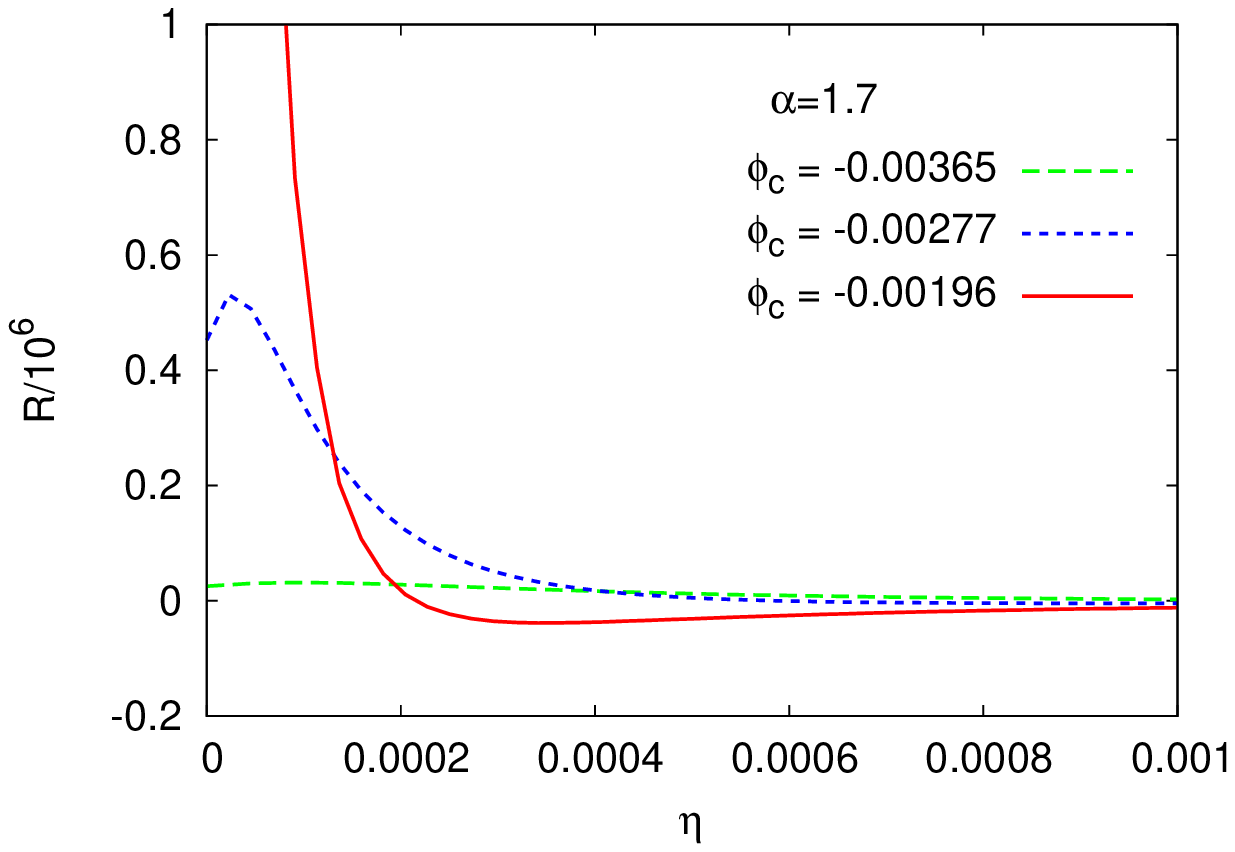}
(d)\includegraphics[width=.47\textwidth, angle =0]{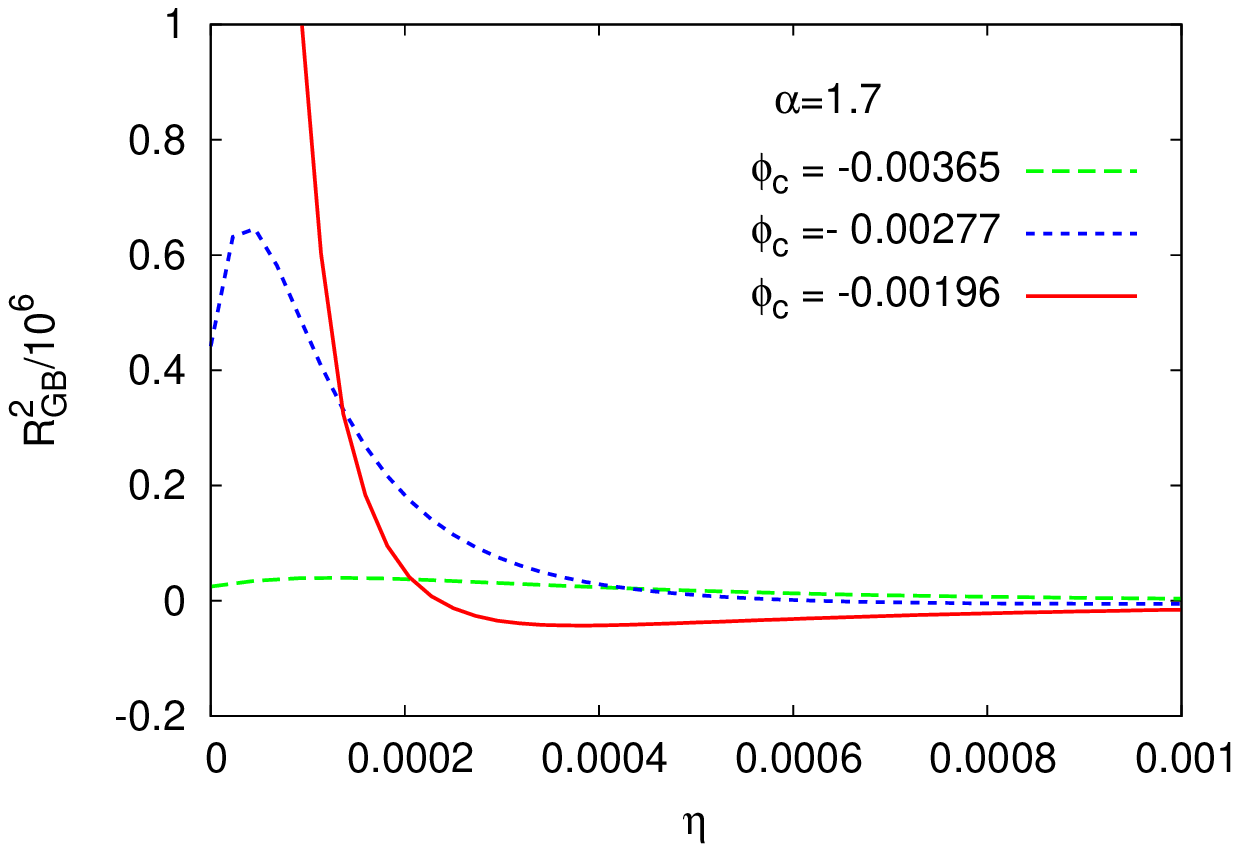}
\\
(e)\includegraphics[width=.47\textwidth, angle =0]{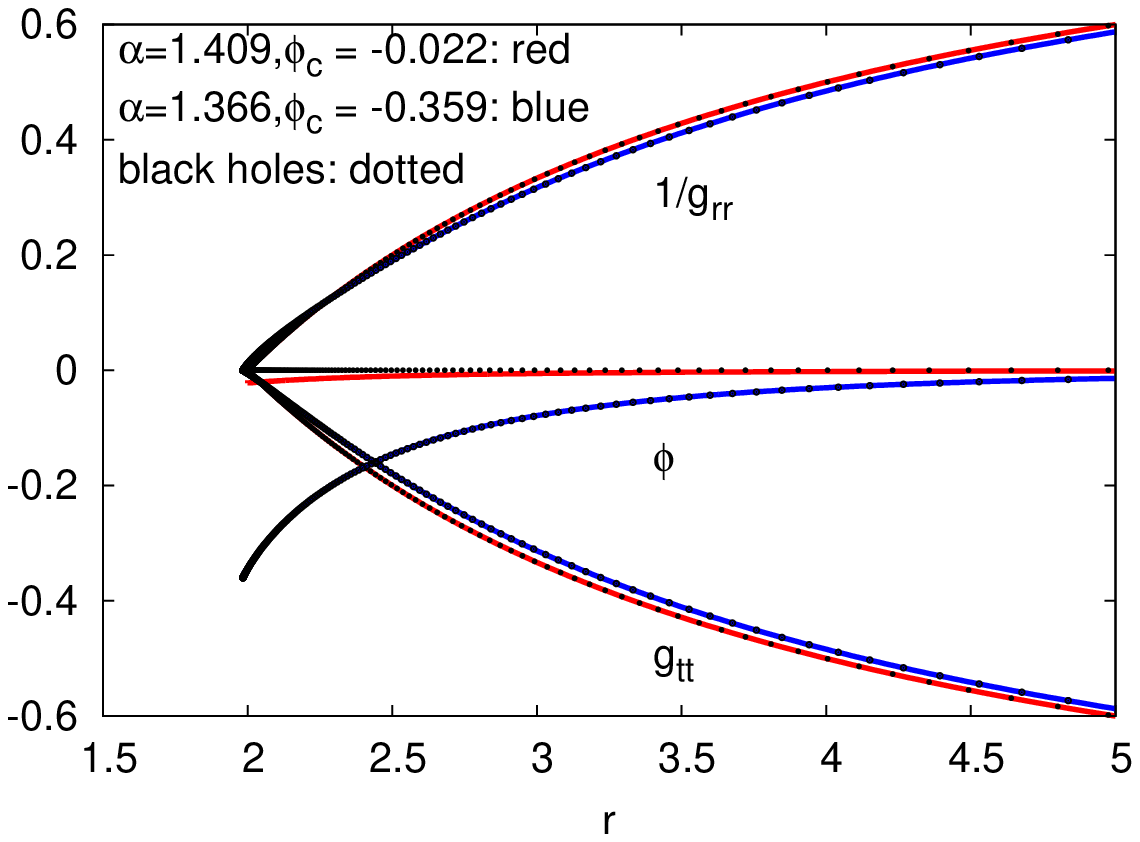}
(f)\includegraphics[width=.47\textwidth, angle =0]{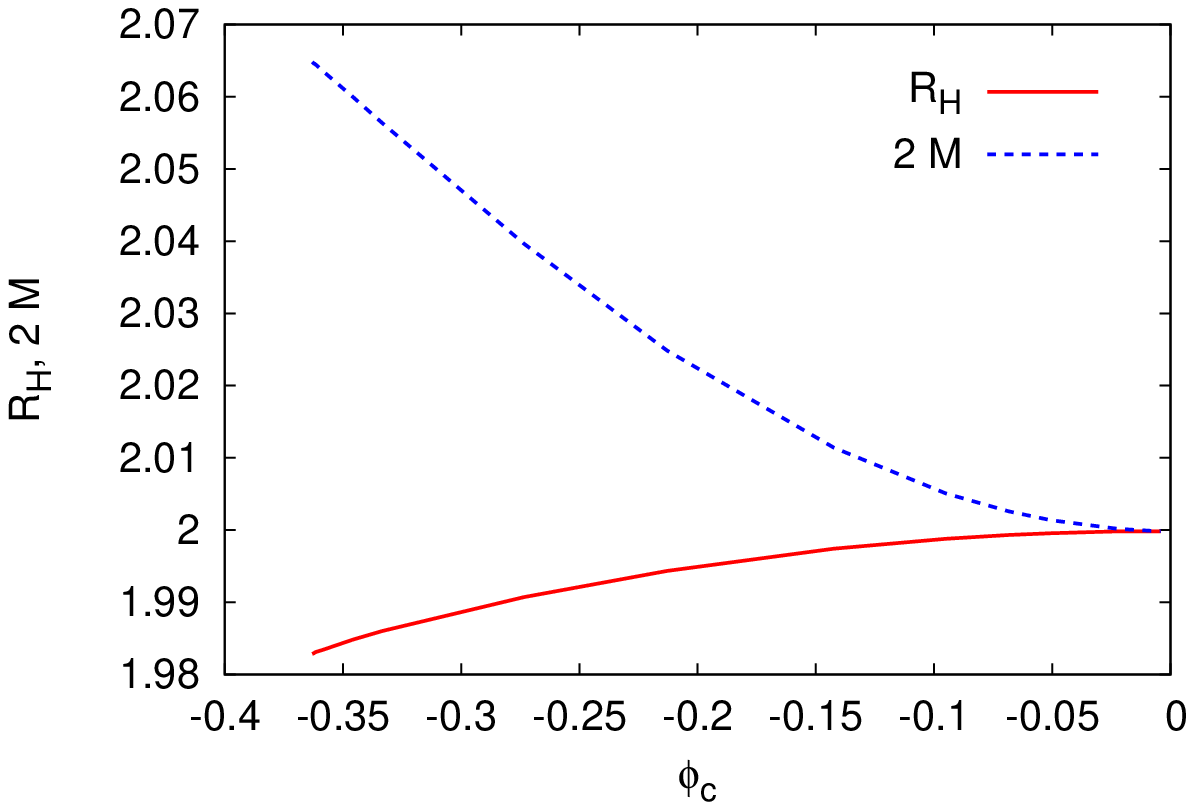}
\end{center}
\caption{Boundary of the domain of existence
(examples with self-interaction):
(a)-(b) convergence towards Scharzschild black hole solutions:
scalar curvature $R$ (a) and Gauss-Bonnet term $R^2_{GB}$ (b)
vs radial coordinate $\eta$  for $\alpha = 1.5$ and several values of $\phi_c$;
(c)-(d) emergence of a curvature singularity at the center:
scalar curvature $R$ (c) and Gauss-Bonnet term $R^2_{GB}$ (d)
vs radial coordinate $\eta$  for $\alpha = 1.1$ and several values of $\phi_c$;
(e) convergence towards scalarized EsGB black hole solutions:
metric functions $g_{tt}$ and $g_{rr}$ of 2 solutions 
vs Schwarzschild radial coordinate $r$ (red: $\alpha=1.409$, 
$\phi_c=-0.022$)
and (blue: $\alpha=1.366$, 
$\phi_c=-0.359$);
(f) set of limiting scalarized (fundamental) EsBG black hole solutions: 
the mass $2M$, and circumferential horizon radius $R_{\text{H}}$ vs the value
of the scalar field at the center $\phi_c$.
}
\label{fig_limits2}
\end{figure}

In Fig.~\ref{fig_scaled}(a) and (b) we exhibit the mass $M$ of the wormhole solutions
versus the circumferentical radius $r_c$ at the center for 
a set of fixed values of the coupling constant $\alpha$,
again with (a) and without (b) self-interaction.
In these diagrams the sets of limiting solutions are clearly visible.
The solid blue dots curve represent the end points
of regular wormhole solutions, where cusp singularities are encountered.
The solid green curve forms the boundary
where singularities at the center arise. 
The solid black curve represents the limit where scalarized EsGB black holes
are encountered, where $r_c$ represents the horizon radius.
We illustrate these limits with examples below.
We have also indicated (dotted black) the line $M=r_c/2$,
corresponding to Schwarzschild black holes with horizon radius $r_c$.

Fig.~\ref{fig_scaled}(c) and (d) are diagrams for
the corresponding scaled quantities. Here the scaled coupling constant
$\alpha/M^2$ is shown versus the scaled circumferentical radius $r_c/M$
at the center, to provide a dimensionless representation of the
domain of existence.
These diagrams again illustrate the huge effect of the self-interaction,
increasing the domain of existence vastly,
in particular, when the scaled quantities are considered.

We now turn to the discussion of the boundary of the domain of existence.
Solutions with cusp singularities form the largest part of the boundary.
The emergence of cusp singularities is related to the determinant
that is encountered upon diagonalization of the set of second order ODEs
with respect to the second derivatives of the functions 
(see also \cite{Antoniou:2019awm,Kleihaus:2019rbg,Kleihaus:2020qwo}).
This determinant may possess a node at some value $\eta_\star$ of the radial coordinate.
Since diagonalization involves devision by the determinant,
the equations of motion are then no longer regular at $\eta_\star$,
but feature a cusp singularity.

We demonstrate how such cusp singularities form in Fig.~\ref{fig_limits}(a) and (b),
where we show for the metric function $f_1$ (a)  and the scalar function $\phi$ (b)
the first derivative $f_{1,\eta}$ and  $\phi_{,\eta}$ versus the radial coordinate $\eta$
for the coupling constant $\alpha=1.0$ and two values
of the scalar field at the center $\phi_c$: 
for $\phi_c=-0.5179$ the second derivatives
$f_{1,\eta\eta}$ and  $\phi_{,\eta\eta}$ develop a jump at some value $\eta_\star$ (blue curves),
whereas for $\phi_c=-0.1612$  the second derivatives
$f_{1,\eta\eta}$ and $\phi_{,\eta\eta}$ diverge at some value $\eta_\star$ (red curves).
The second derivatives $f_{1,\eta\eta}$ and $\phi_{,\eta\eta}$ 
diverge at $\eta_\star$, when the determinant behaves as $(\eta-\eta_\star)^g$ with $g<1$.
In contrast, when the determinant behaves as $(\eta-\eta_\star)$, a jump 
of $f_{1,\eta\eta}$ and $\phi_{,\eta\eta}$ arises.
Associated with these jumps/divergences are divergences of the
curvature scalar $R$ and the GB scalar $R_{\rm GB}$,
as seen in Fig.~\ref{fig_limits}(c) and (d),
where the curvature scalar $R$ (c) and the GB scalar $R_{\rm GB}$ (d)
are shown for the same solutions.



In  Fig.~\ref{fig_limits2}(a)-(d) we consider the part of the boundary, 
where a singularity is approached at the center.
In this case the scalar field $\phi$ goes to zero at the center, $\phi_c \to 0$,  
while its second order derivative diverges to minus infinity at the center, $\phi'(0) \to -\infty$.
At the same time 
the value of the metric function $f_0$ diverges to minus infinity at the center, $f_{0t} \to -\infty$.
Since $\phi_c \to 0$, this part of the boundary resides on the vertical axis
in all plots where $\phi_c$ is shown on the horizontal axis.

For $1.41 \leq \alpha \leq 1.56$ the wormhole solutions (with fixed $\alpha$) converge pointwise to 
the Schwarzschild black hole with unit mass. With decreasing  $|\phi_c|$ 
the deviation from the Schwarzschild black hole decreases on an increasing interval.
In the limit $\phi_c \to 0$ the wormholes coincide with the Schwarzschild black hole
except at the center. The derivative of the scalar field $\phi'(0)$  at the center 
remains finite in this limit, but its second order derivative diverges.
We demonstrate this limiting behaviour for a family
of wormhole solutions with fixed $\alpha=1.5$ and 
$\phi_c = -0.0312$, $-0.0031$ and $-0.0002$. 
In Fig.~\ref{fig_limits2}(a) and (b) we show the  
curvature invariants $R$ (a) and $R^2_{GB}$  (b) versus the radial coordinate
$\eta$ close to the center. The black dots mark the values of the curvature invariants at 
the center. We observe that no curvature singularity emerges.

However, when $\alpha > 1.56$ the limiting behavior changes.
In this case the limit  $\phi_c \to 0$ is characterized by a diverging derivative 
of the scalar field at the center, $\phi'(0) \to -\infty$, and curvature singularities.
This is shown in Fig.~\ref{fig_limits2}(c) and (d) where the  
curvature invariants $R$ (c) and $R^2_{GB}$  (d) are plotted versus the 
radial coordinate $\eta$ for a family of wormhole solutions with fixed value of
$\alpha=1.7$ and  $\phi_c = -0.00365$, $-0.00277$, $-0.00196$.

Interestingly, the singular limits to the Schwarzschild black hole are characterized by a throat at 
the center, whereas the limits to curvature singularities are characterized by an equator.
%

In Fig.~\ref{fig_limits2}(e) we show the metric functions  $g_{tt}$ and $g_{rr}$
and the scalar field function $\phi$
for two (self-interacting) wormhole solutions close to the black hole limit
(with $\alpha=1.409$, $\phi_c=-0.022$ and $\alpha=1.366$, $\phi_c=-0.359$).
For comparison we exhibit the (fundamental) EsGB black hole solutions
for the same values of $\alpha$ and $\phi_c$,
and thus demonstrate the approach of the wormhole solutions to the EsGB black hole boundary.
Fig.~\ref{fig_limits2}(f) shows the mass $2M$, and the circumferential horizon radius $R_{\text{H}}$ 
of the branch of (self-interacting fundamental) EsGB black holes 
versus the value of the scalar field at the center $\phi_c$.
These EsGB black holes branch off from the Schwarzschild solution
at $\alpha=\approx 1.41$, where $\phi_c=0$.

\subsection{Embedding diagrams and energy conditions}

\begin{figure}[t!]
\begin{center}
(a)\includegraphics[width=.45\textwidth, angle =0]{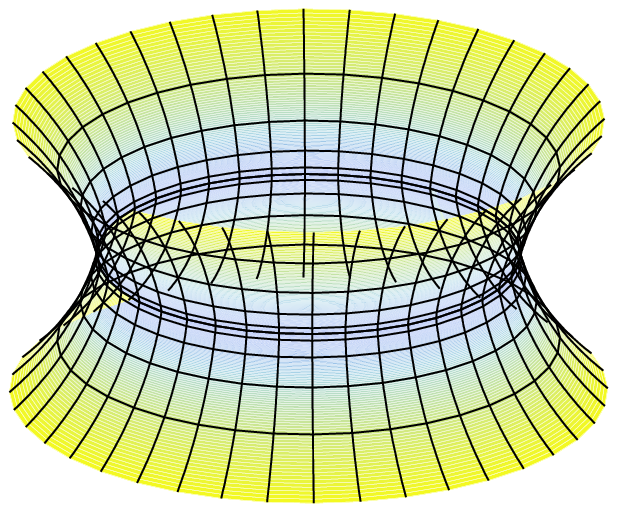}
(b)\includegraphics[width=.45\textwidth, angle =0]{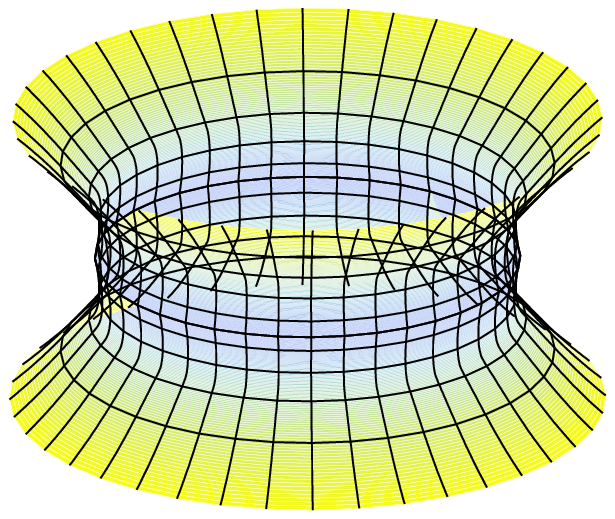}
\end{center}
\caption{Embeddings of the equatorial plane
(examples with self-interaction):
(a) single throat wormhole with coupling constant $\alpha =1.7$ and
value of the scalar field at the center $\phi_c = -0.0121$;
(b) double throat wormhole with $\alpha =1.7$ and $\phi_c = -0.03$.
}
\label{fig_emb}
\end{figure}
\begin{figure}[h!]
\begin{center}
(a)\includegraphics[width=.47\textwidth, angle =0]{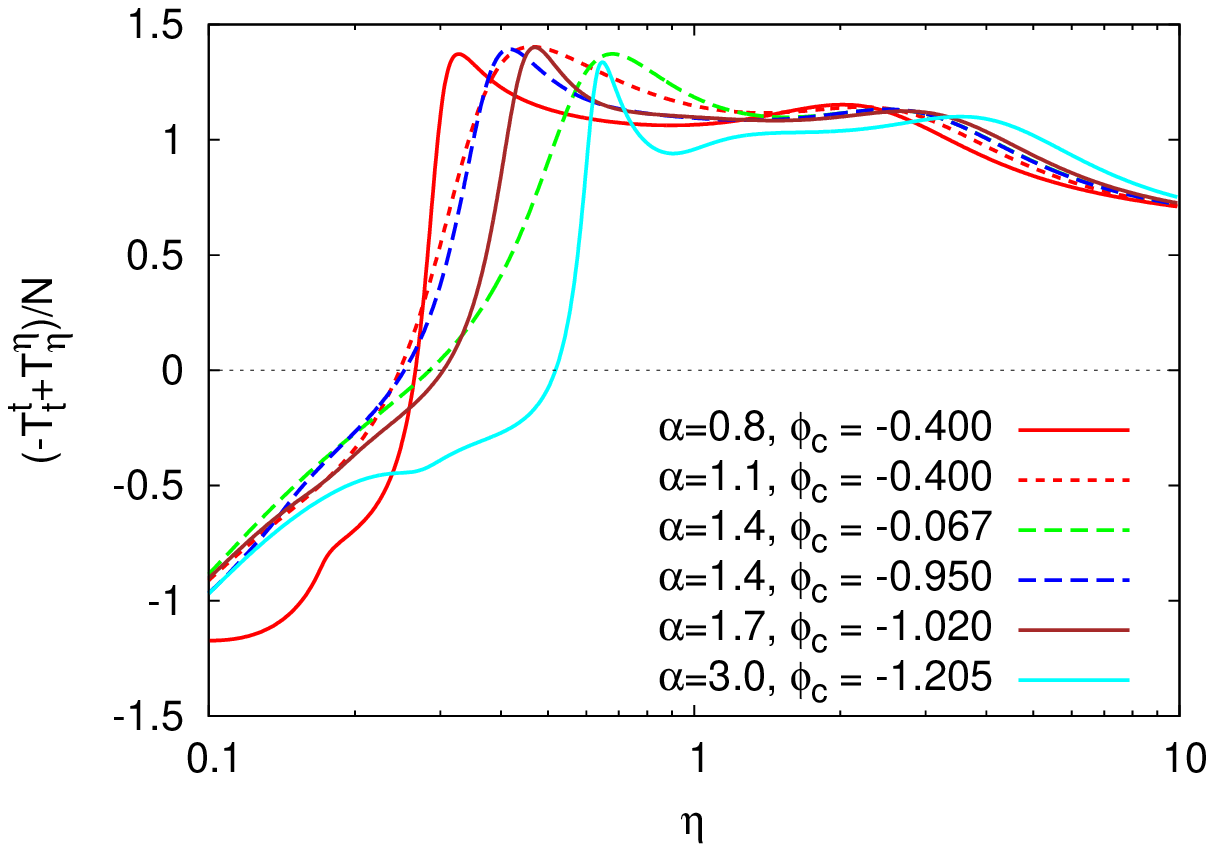}
(b)\includegraphics[width=.47\textwidth, angle =0]{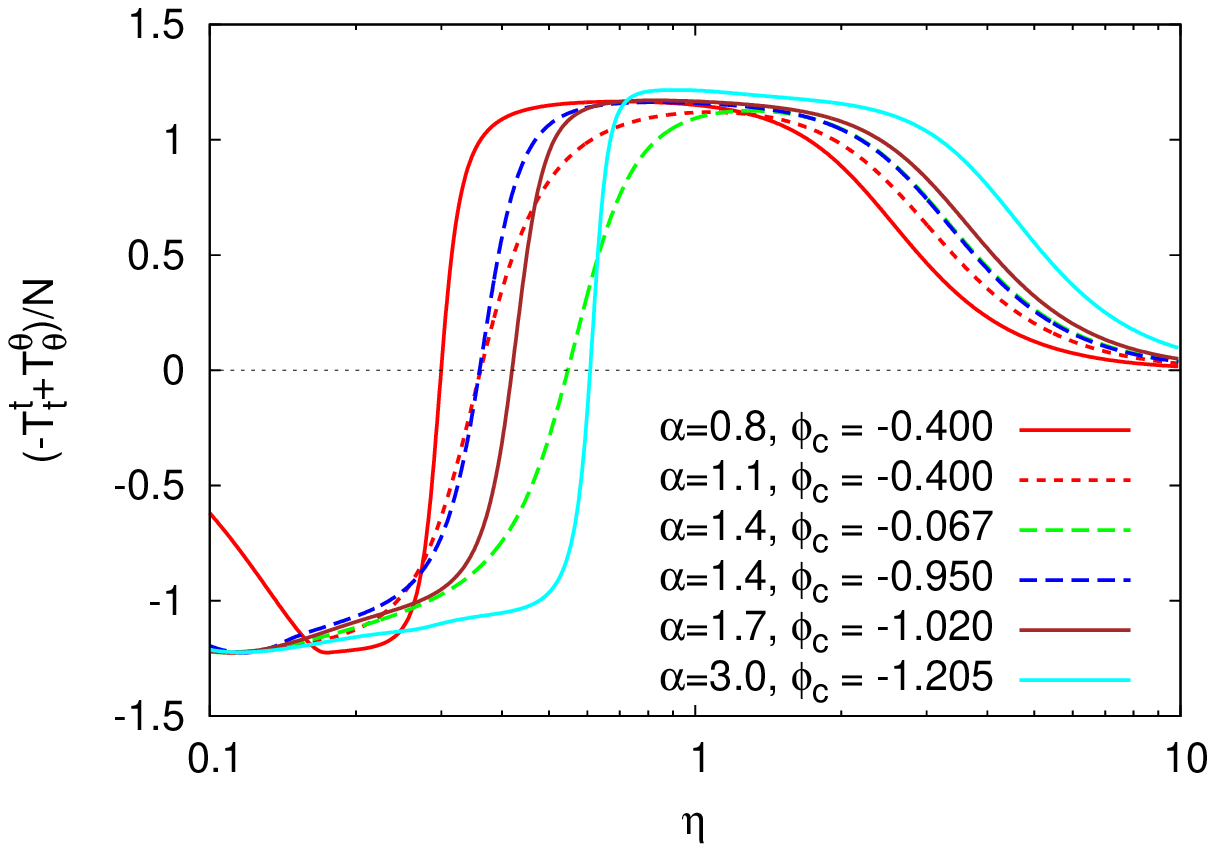}
\end{center}
\caption{Violation of the energy conditions 
(examples with self-interaction):
Combinations of stress-energy tensor
$(- T^t_t+T^\eta_\eta)/N$ (a) and $(- T^t_t+T^\theta_\theta)/N$ (b)
with $N=\sqrt{( T^t_t)^2+(T^r_r)^2+2(T^\theta_\theta)^2}$
vs radial coordinate $\eta$ showing regions of NEC violation
for several values of the coupling constant
$\alpha$ and the scalar field at the center $\phi_c$. 
}
\label{fig_stress}
\end{figure}

We visualize the geometry of a typical single wormhole in Fig.~\ref{fig_emb}(a),
where we have chosen for the coupling constant the value $\alpha =1.7$ and
and for the scalar field at the center $\phi_c = -0.0121$.
Fig.~\ref{fig_emb}(b) shows the geometry of a typical wormhole 
with an equator and a double throat, obtained with the parameter choice
$\alpha =1.7$ and $\phi_c = -0.03$.
The equator and the throats are clearly visible in the figure.
Both solutions belong to the same family ($\alpha =1.7$)
of self-interacting wormholes.
As $\phi_c$ is increased, the single throat (minimum) at the center
first turns into a degenerate throat (saddle point)
at a critical value of $\phi_c$,
and then becomes an equator (maximum),
concealed by a throat (minimum) on each side.

In Fig.~\ref{fig_stress} we exhibit the quantities
$(- T^t_t+T^\eta_\eta)/N$ (a) and $(- T^t_t+T^\theta_\theta)/N$ (b)
with $N=\sqrt{( T^t_t)^2+(T^r_r)^2+2(T^\theta_\theta)^2}$
versus the radial coordinate $\eta$ for several values of the coupling constant
$\alpha$ and  the scalar field at the center $\phi_c$.
$(\alpha \ , \ \phi_c)=$ and  $(0.8,-0.4)$, $(1.1,-0.4)$, $(1.4,-0.067)$, $(1.4,-0.95)$,
$(1.7,-1.02)$, $(3.0,-1.205)$.
When $- T^t_t+T^\eta_\eta$ and $- T^t_t+T^\theta_\theta$ are negative,
the NEC is violated.
Inspection of the figures shows that the NEC is indeed violated,
as it must be.
The violation always occurs in 
the inner regions of the wormhole spacetimes.

\section{Conclusions}

EsGB theories have very attractive properties.
In particular, they allow for several types of interesting compact solutions.
Like GR, they possess black hole solutions
\cite{Antoniou:2017acq,Doneva:2017bvd,Silva:2017uqg,Antoniou:2017hxj,Blazquez-Salcedo:2018jnn,Doneva:2018rou,Minamitsuji:2018xde,Silva:2018qhn,Brihaye:2018grv,
Myung:2018jvi,Bakopoulos:2018nui,Doneva:2019vuh,Myung:2019wvb,Cunha:2019dwb,
Macedo:2019sem,Hod:2019pmb,Bakopoulos:2019tvc, Collodel:2019kkx,Bakopoulos:2020dfg,
Blazquez-Salcedo:2020rhf,Blazquez-Salcedo:2020caw,
Kanti:1995vq,Torii:1996yi,Guo:2008hf,Pani:2009wy,Pani:2011gy,Kleihaus:2011tg,Ayzenberg:2013wua,Ayzenberg:2014aka,Maselli:2015tta,Kleihaus:2014lba,Kleihaus:2015aje,Blazquez-Salcedo:2016enn,Cunha:2016wzk,Zhang:2017unx,Blazquez-Salcedo:2017txk,Konoplya:2019hml,Zinhailo:2019rwd}.
However, these black hole solutions may carry gravitational scalar hair.
Depending on the coupling function to the GB term,
the scalar hair of the black holes can arise due to 
curvature induced spontaneous scalarization.

Besides black holes EsGB theories allow for wormhole solutions,
since their effective stress energy tensor provides for
violation of the energy conditions via the gravitational sector
\cite{Kanti:2011jz,Kanti:2011yv,Antoniou:2019awm}. 
While previously wormhole solutions in EsGB theories
were obtained for massless scalar fields,
we have here considered wormholes in the presence of
a mass term and a sextic self-interaction,
inspired by boson stars.
Moreover, we have employed a type of coupling function
allowing for curvature induced spontaneously scalarized black holes.

We have mapped out the domain of existence of these wormhole solutions,
varying the GB coupling constant $\alpha$.
The boundary of the domain of existence consists mostly
of solutions where a cusp singularity is encountered.
Here the second derivative of some functions either has a jump
or diverges. Consequently, also the curvature scalars diverge.
A small part of the boundary is provided by solutions,
where singularities are encountered at the center
of the configurations. Here the scalar field vanishes at the center,
while its second order derivative diverges along with one of the metric functions.
Some part of this boundary is charcterized by curvature singularities.
The remaining part of the boundary is constituted by the set of 
scalarized EsGB black holes, together with the marginally stable
Schwarzschild black hole.

The domain of existence is significantly increased by the
presence of the self-interaction as compared to a
the case with a mass term only.
This is similar to case of boson stars,
where the sextic self-interaction allows for 
a much larger set of solutions, which, in particular,
possess much higher masses
\cite{Friedberg:1986tq,Kleihaus:2005me}).
Here the sextic self-interaction also leads to more massive
wormholes in the new region in parameter space
available due to the self-interaction.

Most of the wormhole possess a single throat at the center.
However, there is a small region in parameter space,
present already for a mass term only
(and also for vanishing mass),
where the wormholes develop a maximum at the center
surrounded by a minimum on each side.
In this case, the wormhole solutions possess
an equator, that is connected to each asymptotically
flat region via a throat.

At the center, a shell of ordinary matter like, for instance,
dust can be invoked to obtain solutions
that are regular in both asymptotically flat regions
and symmetric with respect to coordinate inflection,
$\eta \to - \eta$. 
The junction conditions \cite{Israel:1966rt,Davis:2002gn}
can be satisfied with ordinary matter,
and there is no need for any type of exotic matter
to obtain regular symmetric wormhole solutions.
This is different from GR, where exotic matter
is needed to obtain the necessary violation of the
energy conditions.

It will be interesting to investigate these wormhole solutions
and their properties further.
Since wormholes represent compact objects,
that can mimick black holes to some extent,
the next objectives will be to study the (possible) lightrings of these objects  
\cite{Cardoso:2014sna,Cunha:2017qtt}
and to investigate the (possible) presence of echoes 
\cite{Cardoso:2016rao,Cardoso:2016oxy,Cardoso:2017cqb}.
These objectives have recently already been addressed for
a further type of compact solutions,
that EsGB theories allow for, 
namely particle-like solutions
\cite{Kleihaus:2019rbg,Kleihaus:2020qwo}.
It should certainly be interesting to also extend these particle-like solutions
to the case of a non-vanishing scalar field potential,
as has been done here for the wormhole solutions \cite{Antoniou:2019awm}.

\section*{Acknowledgement}

BK and JK gratefully acknowledge support by the
DFG Research Training Group 1620 {\sl Models of Gravity}
and the COST Action CA16104. 

\section{Appendix}


\begin{thebibliography}{unsrtnat}

\bibitem{Morris:1988cz}
  M.~S.~Morris and K.~S.~Thorne,
  Am.\ J.\ Phys.\  {\bf 56} (1988) 395.

\bibitem{Visser:1995cc} 
  M.~Visser,
  ``Lorentzian wormholes: From Einstein to Hawking,''
  Woodbury, USA: AIP (1995) 

\bibitem{Lobo:2017eum}
F.~S.~Lobo,
Fundam. Theor. Phys. \textbf{189} (2017) 1.

\bibitem{Hochberg:1990is}
  D.~Hochberg,
  Phys.\ Lett.\  {\bf B251}, 349 (1990).

\bibitem{Fukutaka:1989zb}
  H.~Fukutaka, K.~Tanaka, K.~Ghoroku,
  Phys.\ Lett.\  {\bf B222}, 191 (1989).

\bibitem{Ghoroku:1992tz}
  K.~Ghoroku, T.~Soma,
  Phys.\ Rev.\  {\bf D46}, 1507 (1992).

\bibitem{Furey:2004rq}
  N.~Furey, A.~DeBenedictis,
  Class.\ Quant.\ Grav.\  {\bf 22}, 313 (2005).

\bibitem{Bronnikov:2009az}
  K.~A.~Bronnikov and E.~Elizalde,
  Phys.\ Rev.\  D {\bf 81}, 044032 (2010).


\bibitem{Kanti:2011jz} 
  P.~Kanti, B.~Kleihaus and J.~Kunz,
  Phys.\ Rev.\ Lett.\  {\bf 107}, 271101 (2011).

\bibitem{Kanti:2011yv}
  P.~Kanti, B.~Kleihaus and J.~Kunz,
  Phys.\ Rev.\ D {\bf 85}, 044007 (2012).
%
  
\bibitem{Lobo:2009ip} 
  F.~S.~N.~Lobo and M.~A.~Oliveira,
  Phys.\ Rev.\ D {\bf 80}, 104012 (2009).

\bibitem{Harko:2013yb} 
  T.~Harko, F.~S.~N.~Lobo, M.~K.~Mak and S.~V.~Sushkov,
  Phys.\ Rev.\ D {\bf 87}, 067504 (2013).

\bibitem{Zwiebach:1985uq}
  B.~Zwiebach,
  Phys.\ Lett.\  {\bf 156B}, 315 (1985).

\bibitem{Gross:1986mw}
  D.~J.~Gross and J.~H.~Sloan,
  Nucl.\ Phys.\ B {\bf 291}, 41 (1987).

\bibitem{Metsaev:1987zx}
  R.~R.~Metsaev, A.~A.~Tseytlin,
  Nucl.\ Phys.\  {\bf B293 }, 385 (1987).

\bibitem{Sotiriou:2013qea} 
  T.~P.~Sotiriou and S.~Y.~Zhou,
  Phys.\ Rev.\ Lett.\  {\bf 112}, 251102 (2014)

\bibitem{Sotiriou:2014pfa} 
  T.~P.~Sotiriou and S.~Y.~Zhou,
  Phys.\ Rev.\ D {\bf 90}, 124063 (2014)

\bibitem{Antoniou:2017acq}
G.~Antoniou, A.~Bakopoulos and P.~Kanti,
Phys.\ Rev.\ Lett.\  {\bf 120}, no. 13, 131102 (2018);
Phys.\ Rev.\ D {\bf 97} (2018) no.8,  084037.

\bibitem{Doneva:2017bvd}
D.~D.~Doneva and S.~S.~Yazadjiev,
Phys.\ Rev.\ Lett.\  {\bf 120}, no. 13, 131103 (2018).

\bibitem{Silva:2017uqg}
H.~O.~Silva, J.~Sakstein, L.~Gualtieri, T.~P.~Sotiriou and E.~Berti,
Phys.\ Rev.\ Lett.\  {\bf 120}, no. 13, 131104 (2018).


\bibitem{Horndeski:1974wa} 
  G.~W.~Horndeski,
  Int.\ J.\ Theor.\ Phys.\  {\bf 10}, 363 (1974).

\bibitem{Charmousis:2011bf} 
  C.~Charmousis, E.~J.~Copeland, A.~Padilla and P.~M.~Saffin,
  Phys.\ Rev.\ Lett.\  {\bf 108}, 051101 (2012)

\bibitem{Kobayashi:2011nu} 
  T.~Kobayashi, M.~Yamaguchi and J.~Yokoyama,
  Prog.\ Theor.\ Phys.\  {\bf 126}, 511 (2011)

\bibitem{Israel:1966rt} 
  W.~Israel,
  Nuovo Cim.\ B {\bf 44S10} (1966) 1
   [Nuovo Cim.\ B {\bf 44} (1966) 1]
   Erratum: [Nuovo Cim.\ B {\bf 48} (1967) 463].

\bibitem{Davis:2002gn} 
  Phys.\ Rev.\ D {\bf 67} (2003) 024030.

\bibitem{Kanti:1995vq}
  P.~Kanti, N.~E.~Mavromatos, J.~Rizos, K.~Tamvakis and E.~Winstanley,
  Phys.\ Rev.\  D {\bf 54} (1996) 5049.

\bibitem{Antoniou:2019awm}
G.~Antoniou, A.~Bakopoulos, P.~Kanti, B.~Kleihaus and J.~Kunz,
Phys. Rev. D \textbf{101} (2020) no.2, 024033


\bibitem{Antoniou:2017hxj} 
  G.~Antoniou, A.~Bakopoulos and P.~Kanti,
  Phys.\ Rev.\ D {\bf 97}, no. 8, 084037 (2018)

\bibitem{Blazquez-Salcedo:2018jnn} 
  J.~L.~Bl\'{a}zquez-Salcedo, D.~D.~Doneva, J.~Kunz and S.~S.~Yazadjiev,
  Phys.\ Rev.\ D {\bf 98}, no. 8, 084011 (2018).

\bibitem{Doneva:2018rou}
D.~D.~Doneva, S.~Kiorpelidi, P.~G.~Nedkova, E.~Papantonopoulos and S.~S.~Yazadjiev,
Phys.\ Rev.\ D \textbf{98}, no.10, 104056 (2018)

\bibitem{Minamitsuji:2018xde} 
  M.~Minamitsuji and T.~Ikeda,
  Phys.\ Rev.\ D {\bf 99}, no. 4, 044017 (2019)

\bibitem{Silva:2018qhn} 
  H.~O.~Silva, C.~F.~B.~Macedo, T.~P.~Sotiriou, L.~Gualtieri, J.~Sakstein and E.~Berti,
  Phys.\ Rev.\ D {\bf 99}, no. 6, 064011 (2019)

\bibitem{Brihaye:2018grv} 
  Y.~Brihaye and L.~Ducobu,
  Phys.\ Lett.\ B {\bf 795}, 135 (2019)

\bibitem{Myung:2018jvi}
Y.~S.~Myung and D.~Zou,
Phys.\ Lett.\ B \textbf{790}, 400-407 (2019)

\bibitem{Bakopoulos:2018nui}
 A.~Bakopoulos, G.~Antoniou and P.~Kanti,
  Phys.\ Rev.\ D {\bf 99}, no.6,  064003 (2019)

\bibitem{Doneva:2019vuh} 
  D.~D.~Doneva, K.~V.~Staykov and S.~S.~Yazadjiev,
  Phys.\ Rev.\ D {\bf 99}, no. 10, 104045 (2019)

\bibitem{Myung:2019wvb} 
  Y.~S.~Myung and D.~C.~Zou,
  Int.\ J.\ Mod.\ Phys.\ D {\bf 28}, no. 09, 1950114 (2019)

\bibitem{Macedo:2019sem} 
  C.~F.~B.~Macedo, J.~Sakstein, E.~Berti, L.~Gualtieri, H.~O.~Silva and T.~P.~Sotiriou,
  Phys.\ Rev.\ D {\bf 99}, no. 10, 104041 (2019)

\bibitem{Cunha:2019dwb} 
  P.~V.~P.~Cunha, C.~A.~R.~Herdeiro and E.~Radu,
  Phys.\ Rev.\ Lett.\  {\bf 123}, no. 1, 011101 (2019)

\bibitem{Bakopoulos:2019tvc}
  A.~Bakopoulos, P.~Kanti and N.~Pappas,
  Phys.\ Rev.\ D {\bf 101}, no.4,  044026 (2020)

\bibitem{Hod:2019pmb}
  S.~Hod,
  Phys.\ Rev.\ D {\bf 100} (2019) no.6,  064039

\bibitem{Collodel:2019kkx}
L.~G.~Collodel, B.~Kleihaus, J.~Kunz and E.~Berti,
Class.\ Quant.\ Grav.\  \textbf{37}, no.7, 075018 (2020)

\bibitem{Bakopoulos:2020dfg}
A.~Bakopoulos, P.~Kanti and N.~Pappas,
Phys.\ Rev.\ D {\bf 101}, no.8,  084059 (2020)

\bibitem{Blazquez-Salcedo:2020rhf}
J.~L.~Bl\'azquez-Salcedo, D.~D.~Doneva, S.~Kahlen, J.~Kunz, P.~Nedkova and S.~S.~Yazadjiev,
Phys. Rev. D \textbf{101} (2020) no.10, 104006
Antoniou:2019awm
\bibitem{Blazquez-Salcedo:2020caw}
J.~L.~Bl\'azquez-Salcedo, D.~D.~Doneva, S.~Kahlen, J.~Kunz, P.~Nedkova and S.~S.~Yazadjiev,
[arXiv:2006.06006 [gr-qc]].

\bibitem{Torii:1996yi}
T.~Torii, H.~Yajima and K.~i.~Maeda,
Phys.\ Rev.\ D {\bf 55}, 739 (1997)

\bibitem{Guo:2008hf}
  Z.~K.~Guo, N.~Ohta and T.~Torii,
  Prog.\ Theor.\ Phys.\  {\bf 120}, 581 (2008)

\bibitem{Pani:2009wy}
P.~Pani and V.~Cardoso,
Phys.\ Rev.\ D {\bf 79}, 084031 (2009)

\bibitem{Pani:2011gy}
  P.~Pani, C.~F.~B.~Macedo, L.~C.~B.~Crispino and V.~Cardoso,
  Phys.\ Rev.\ D {\bf 84}, 087501 (2011)

\bibitem{Kleihaus:2011tg}
  B.~Kleihaus, J.~Kunz and E.~Radu,
  Phys.\ Rev.\ Lett.\  {\bf 106} (2011) 151104.

\bibitem{Ayzenberg:2013wua}
  D.~Ayzenberg, K.~Yagi and N.~Yunes,
  Phys.\ Rev.\ D {\bf 89}, no. 4, 044023 (2014)

\bibitem{Ayzenberg:2014aka}
D.~Ayzenberg and N.~Yunes,
Phys.\ Rev.\ D {\bf 90}, 044066 (2014)

\bibitem{Maselli:2015tta}
A.~Maselli, P.~Pani, L.~Gualtieri and V.~Ferrari,
Phys.\ Rev.\ D {\bf 92}, no. 8, 083014 (2015)

\bibitem{Kleihaus:2014lba}
  B.~Kleihaus, J.~Kunz and S.~Mojica,
  Phys.\ Rev.\ D {\bf 90}, no. 6, 061501 (2014)

\bibitem{Kleihaus:2015aje}
  B.~Kleihaus, J.~Kunz, S.~Mojica and E.~Radu,
  Phys.\ Rev.\ D {\bf 93}, no. 4, 044047 (2016)

\bibitem{Blazquez-Salcedo:2016enn} 
  J.~L.~Bl\'{a}zquez-Salcedo, C.~F.~B.~Macedo, V.~Cardoso, V.~Ferrari, L.~Gualtieri, F.~S.~Khoo, J.~Kunz and P.~Pani,
  Phys.\ Rev.\ D {\bf 94}, no. 10, 104024 (2016)

\bibitem{Cunha:2016wzk} 
  P.~V.~P.~Cunha, C.~A.~R.~Herdeiro, B.~Kleihaus, J.~Kunz and E.~Radu,
  Phys.\ Lett.\ B {\bf 768}, 373 (2017)

\bibitem{Zhang:2017unx} 
  H.~Zhang, M.~Zhou, C.~Bambi, B.~Kleihaus, J.~Kunz and E.~Radu,
  Phys.\ Rev.\ D {\bf 95}, no. 10, 104043 (2017)

\bibitem{Blazquez-Salcedo:2017txk}
J.~L.~Bl\'azquez-Salcedo, F.~S.~Khoo and J.~Kunz,
Phys.\ Rev.\ D \textbf{96}, no.6, 064008 (2017)

\bibitem{Konoplya:2019hml}
R.~Konoplya, A.~Zinhailo and Z.~Stuchlík,
Phys.\ Rev.\ D \textbf{99}, no.12, 124042 (2019)

\bibitem{Zinhailo:2019rwd}
A.~Zinhailo,
Eur.\ Phys.\ J.\ C \textbf{79}, no.11, 912 (2019)

\bibitem{Friedberg:1986tq}
R.~Friedberg, T.~Lee and Y.~Pang,
Phys. Rev. D \textbf{35}, 3658 (1987)

\bibitem{Kleihaus:2005me}
B.~Kleihaus, J.~Kunz and M.~List,
Phys. Rev. D \textbf{72}, 064002 (2005)

\bibitem{Brihaye:2020dgo}
Y.~Brihaye and J.~Renaux,
[arXiv:2004.12138 [gr-qc]].

\bibitem{Ascher:1979iha}
U.~Ascher, J.~Christiansen and R.~Russell,
Math. Comput. \textbf{33} (1979) no.146, 659-679

\bibitem{Kleihaus:2019rbg}
B.~Kleihaus, J.~Kunz and P.~Kanti,
Phys. Lett. B \textbf{804}, 135401 (2020)

\bibitem{Kleihaus:2020qwo}
B.~Kleihaus, J.~Kunz and P.~Kanti,
[arXiv:2005.07650 [gr-qc]].

\bibitem{Cardoso:2014sna} 
  V.~Cardoso, L.~C.~B.~Crispino, C.~F.~B.~Macedo, H.~Okawa and P.~Pani,
  Phys.\ Rev.\ D {\bf 90}, no. 4, 044069 (2014)

\bibitem{Cunha:2017qtt} 
  P.~V.~P.~Cunha, E.~Berti and C.~A.~R.~Herdeiro,
  Phys.\ Rev.\ Lett.\  {\bf 119}, no. 25, 251102 (2017)

\bibitem{Cardoso:2016rao} 
  V.~Cardoso, E.~Franzin and P.~Pani,
  Phys.\ Rev.\ Lett.\  {\bf 116}, no. 17, 171101 (2016)
  Erratum: [Phys.\ Rev.\ Lett.\  {\bf 117}, no. 8, 089902 (2016)]

\bibitem{Cardoso:2016oxy} 
  V.~Cardoso, S.~Hopper, C.~F.~B.~Macedo, C.~Palenzuela and P.~Pani,
  Phys.\ Rev.\ D {\bf 94}, no. 8, 084031 (2016).

\bibitem{Cardoso:2017cqb} 
  V.~Cardoso and P.~Pani,
  Nat.\ Astron.\  {\bf 1}, no. 9, 586 (2017)



\end{thebibliography}
\end{document}